\documentclass[aps,prd,onecolumn,nofootinbib,superscriptaddress]{revtex4}
\usepackage{float}
\usepackage{graphicx}
\usepackage{amsmath}
\usepackage{amsfonts}
\usepackage{amssymb,ulem}
\usepackage{color}%
\usepackage{dcolumn}
\usepackage{subfigure}

\usepackage{MnSymbol,wasysym}
\usepackage{braket,diagbox}
\usepackage{eurosym}
\usepackage{calrsfs}
\usepackage[usenames,dvipsnames,svgnames]{xcolor}

\newcommand{\RNum}[1]{\uppercase\expandafter{\romannumeral #1\relax}}
\usepackage[colorlinks=true,linkcolor=blue,urlcolor=blue,filecolor=black,citecolor=cyan]{hyperref}

\begin{document}


\title{Strong gravitational lensing and shadow constraint from M87* of slowly rotating Kerr-like black hole}
\author{Xiao-Mei Kuang}
\email{xmeikuang@yzu.edu.cn}
\affiliation{Center for Gravitation and Cosmology, College of Physical Science and Technology, Yangzhou University, Yangzhou, 225009, China}

\author{Ali \"Ovg\"un}
\email{ali.ovgun@emu.edu.tr}
\affiliation{Physics Department, Eastern Mediterranean University, Famagusta, 99628 North
Cyprus via Mersin 10, Turkey.}

\date{\today}

\begin{abstract}

Motivated by (i) more and more interest in strong gravitational lensing by supermassive black holes due to the achievement of EHT observations, (ii) the ongoing popular topic on the possibility of Lorentz symmetry  being broken in gravitation and its consequences,
 we will apply the Einstein bumblebee gravity with Lorentz violation (LV) to the study of strong gravitational lensing effect and the black hole shadow of slowly rotating Kerr-like black hole. In the strong gravitational lensing sector, we first calculate the deflection angel; then treating the slowly rotating Kerr-like black hole  as supermassive M87* black hole, we evaluate the gravitational lensing observables (position, separation and magnification) and the time delays between the relativistic images. In the black hole shadow sector, we show the effect of LV parameter on the luminosity of the black hole shadow and photon sphere using the infalling spherical accretion. Moreover, we explore the dependence of various shadow observables on the LV parameter, and then give the possible constrain on the LV parameter by M87* black hole of EHT observations. We find that the LV parameter show significant effect on the strong gravitational lensing effect, the black hole shadow and photon sphere luminosity by accretion material.  Our results point out that the future generations of EHT observation may help to distinguish the Einstein bumblebee gravity from GR, and also give a possible constrain on the LV parameter.

\end{abstract}


\maketitle
\tableofcontents
\section{Introduction}
Over the centuries there have been many theoretical physics predictions that have rocked our understanding of how the universe works. One of them is the theory of gravitation, general relativity (GR) developed by Albert Einstein, which stands as one of the best-tested theories in science. One of the basic assumptions of GR
is that of Lorentz invariance, a fundamental symmetry which has been verified to a very high
precision. A fruitful outcome of the Einstein's gravity theory is the black hole that are the most compact objects in the universe, providing probes of the strongest-possible gravitational fields. First, Schwarzschild had provided the solution of general relativity that characterized black holes in the year 1916. However, before the Schwarzschild, Laplace and Michell had considered objects with strong gravitational fields that could trap light using the Newton's gravity theory. Since then existence of the black holes are confirmed by various astrophysical observations; such as detection of gravitational waves (GWs) by LIGO/ VIRGO observations \cite{GW1,GW2,GW3} and the detection of the shadow of the supermassive M87* ans Sgr A* black holes by the Event Horizon Telescope (EHT) \cite{EventHorizonTelescope:2019dse,EventHorizonTelescope:2019ths,EventHorizonTelescope:2019pgp,EventHorizonTelescope:2022xnr,EventHorizonTelescope:2022xqj}.  \\

Rapidly advancing technologies and cross-disciplinary partnerships are accelerating the gravitational waves experiments and shadow observations with some constraints developed based on these findings already being tested in literatures, see \cite{Cunha:2018acu,Perlick:2021aok} for review. The next decade promises further progress in understanding the event horizon of the black hole. Curvature of the spacetime grows so extreme near a black hole's center that Einstein's equations break so that black holes are the good place to search quantum origin of gravity.  \\

In physics, field theories, mathematical descriptions of how field values change in spacetime, for example, GR and Quantum Field Theory (QFT) (the framework underlying elementary particle physics). Although there are many different attempts to find quantum gravity, until now, it is still unclear how GR and QFT should be unified into a consistent theory at the Planck scale ( $10^{19}$ GeV). The effects of the Lorentz symmetry breaking (LSB) at a low energy levels can be observed in experiments in current energy level to see relic of quantum gravity. For this purpose, in 1989 Kostelecky and Samuel had pioneeringly constructed a simple model for spontaneous Lorentz violating (LV) today known as bumblebee gravity \cite{Kostelecky:1989jw} where bumblebee field with a vacuum expectation value causes spontaneous breaks in Lorentz symmetry and the LV appears from the dynamics of a single vector (bumblebee field) $B_\mu$ \cite{Kostelecky:2000mm,Kostelecky:2002ca,Bertolami:2003qs,Cambiaso:2012vb}.  \\

The bumblebee gravity theory has attracted plenty of interest over the years,
especially the construction of black hole solution in such a mode since the black holes  give a large amount of information about the quantum gravity area. Soon in Einstein bumblebee gravity model the exact static black hole, Schwarzschild like solution and its classical test was exhibited  in
\cite{Bertolami:2005bh, Casana:2017jkc}, of which the gravitational deflection angle of light \cite{Ovgun:2018ran,Li:2020wvn}, the Hawking radiation \cite{Kanzi:2019gtu} and the accretion \cite{Yang:2018zef} have been studied. \"Ovg\"un et al. have constructed exact traversable wormhole solution in bumblebee gravity \cite{Ovgun:2018xys} and then Oliveira et al. have studied its  quasinormal modes \cite{Oliveira:2018oha}. The derivation of the exact black hole spacetime metric in the Einstein Hilbert Bumblebee gravity around global monopole field is given in \cite{Gullu:2020qzu}. Santos et al. have studied Gödel solution in the bumblebee gravity \cite{Santos:2014nxm}. Radiative corrections in metric-affine bumblebee model have been studied in \cite{Delhom:2020gfv}.  Later, considering that the rotating black hole is more relative  subsets for astrophysics, an exact Kerr-like metric in Einstein bumblebee gravity have been found in \cite{Ding:2019mal} and more physics, including shadow and gravitational deflection, on this Kerr-like metric have been investigated in \cite{Ding:2019mal,Li:2020dln,DCarvalho:2021zpf}. However, it was soon pointed out in \cite{Ding:2020kfr,Maluf:2022knd,Kanzi:2022vhp} that the Kerr-like metric doesn't satisfy the bumblebee field motion equation, unless the slowly rotating limit (the spin parameter $a^2\to 0$)  is  considered.  \\

As a powerful astrophysical and
cosmological tool,  the gravitational lensing  has been used to test the features of the strong field regime of gravity \cite{Bekenstein:1993fs,Eiroa:2005ag,Sarkar:2006ry,Chen:2009eu,Kumar:2020sag,Islam:2021dyk,Cunha:2015yba,EventHorizonTelescope:2021dqv,Cunha:2016bjh,Junior:2021dyw,Virbhadra:2007kw,Keeton:2005jd,Virbhadra:2008ws,Virbhadra:1998dy,Virbhadra:1999nm,Ovgun:2018fnk,Stefanov:2010xz,Gyulchev:2006zg,Gyulchev:2008ff,Hoekstra:2013via,Javed:2019kon,Javed:2019ynm,Ghosh:2020spb,Uniyal:2022vdu,Pantig:2022toh}.  Among those, the gravitational lensing in the strong gravity regime of black hole have attracted considerable attention since one could get the properties of black hole from it.
First, the lens equation has been introduced by Virbhadra and Ellis for asymptotically flat spacetimes \cite{Virbhadra:1999nm}, then Bozza has derived an analytical method for strong field lensing \cite{Bozza:2002zj} and then Perlick calculated the lensing observables \cite{Perlick:2003vg}, Later Bozza’s technique
was applied to spacetime of charged black hole by Eiroa et al. \cite{Eiroa:2002mk}. Tsukamoto has improved the strong lensing for a general asymptotically flat, static, spherically symmetric spacetime as well as in ultrastatic spacetimes \cite{Tsukamoto:2016jzh}.  \\

In particular, some quantitative observables of strong gravitational lensing effect of Kerr black hole have been managed to be calculated in \cite{Bozza:2002zj,Vazquez:2003zm,Bozza:2003cp}, which inspired plenty of extension, especially in rotating black holes for observational advantage, see for examples \cite{Beckwith:2004ae,Wei:2011nj,Gralla:2019drh,Hsiao:2019ohy}. More recently, motivated by the observations of the supermassive black hole M87* captured by EHT, the authors of \cite{Islam:2021dyk} considered the supermassive black hole M87* as lens, and studied the observables, including deflection angle, the positions, separation, magnification and time delay of relativistic images by  M87*. They found that comparing the Kerr case in GR, the hair of the hairy Kerr black hole have significant effect on the strong gravitational lensing.

On the other hand, the shadow of M87* from EHT observation \cite{EventHorizonTelescope:2019dse,EventHorizonTelescope:2019ths,EventHorizonTelescope:2019pgp} gives us certain constraint on its observables, such as a deviation from circularity $\Delta C \lesssim 0.1$, a axis ratio $1< D_x \lesssim 4/3$ and the angular diameter $\theta_d=42\pm 3\mu as$. Though these constraints on the shadow observables fulfill those for Kerr black hole predicted from GR, the non-Kerr black holes in GR or the (non-)Kerr in modified theories of gravity (MoG) can not be ruled out. Therefore, the EHT observations of shadow and thin accretion disk are then applied as an important tool to constrain the black hole parameters in MoG, and even to distinguish different theories of gravity
\cite{Ayzenberg:2016ynm,Afrin:2021imp, Kumar:2019pjp, Jha:2021bue,  Khodadi:2021gbc, Afrin:2021wlj, Cunha:2019ikd, Khodadi:2020jij, Hioki:2009na, Kumar:2018ple, Tsupko:2017rdo, Banerjee:2019nnj, Kumar:2020owy, Nampalliwar:2020asd, Guo:2020blq,Zeng:2020dco,Cai:2021uov,Wielgus:2021peu, He:2022yse, Dokuchaev:2020wqk,Cimdiker:2021cpz,Ovgun:2018tua,Ovgun:2020gjz,Vagnozzi:2022moj, Roy:2021uye, Vagnozzi:2019apd, Allahyari:2019jqz, Bambi:2019tjh,Lee:2021sws,Brahma:2020eos,Meng:2022kjs,Kuang:2022ojj,Chen:2022lct,Wang:2022kvg,Tang:2022hsu,Abdujabbarov:2016hnw,Abdujabbarov:2015xqa}.  \\

Main aim of this paper is to apply the Einstein bumblebee gravity to the
study of strong gravitational lensing effect and shadow of slowly rotating Kerr-like black hole which satisfies the equation of motion of the model. The significance of our study could stem from two aspects. One is that a flurry of interest in strong gravitational lensing by supermassive black hole has being prompted due to the achievement of EHT observations. The other is that the possibility of Lorentz symmetry  being broken in gravitation and its consequences is ongoing popular topic on debate.
By presupposing the rotating black hole with LV as M87* black hole, we first evaluate the effect of LV parameter on the gravitational lensing observables as well as the time delay between the relativistic images, then we tend to constrain the LV parameter with the use of the shadow observables from EHT observation.  \\

The remaining part of this paper is organized as follows. In Section \ref{sec:background}, we briefly review the Einstein bumblebee gravity and its slowly rotating black hole solution. In Section \ref{sec:SGL}, starting from the null geodesic in the equatorial plane, we evaluate the deflection angle of strong gravitational effect from the black hole, and then by treating the current black hole as M87*, we evaluate the gravitational lensing observables as well as the time delay between the relativistic images and we analyze the effect of the LV parameter. In Section \ref{sec:shadow}, we investigate the black hole shadow cast, photon sphere using the infalling spherical accretion and the shadow observables, and then give the possible constrain on the LV parameter by M87* black hole of EHT observations. We conclude our findings in Section \ref{sec:conclusion}. In this paper, we will use the natural units unless we reassign, and the metric signature $(-,+,+,+)$.  \\

\section{Slowly rotating black hole in Einstein bumblebee gravity}\label{sec:background}
We  will briefly review the slowly rotating Kerr-like black hole in Einstein bumblebee gravity. The bumblebee model is a Standard Model Extension (SME) that violates the Lorentz symmetry through a vector bumblebee field, of which the nonvanishing vacuum expectation value  causes the Lorentz and diffeomorphism violation \cite{Kostelecky:1989jw,Kostelecky:2000mm,Kostelecky:2002ca,Bertolami:2003qs,Cambiaso:2012vb,Bluhm:2004ep}. Moreover, it is also discussed in \cite{Paramos:2014mda,Ding:2015kba,Ding:2016wcf} that this
model is a subset of Einstein-aether theory.
The action of the Einstein bumblebee model is given by
\begin{equation}
S_B =\int d^4y\sqrt{-g}\left[\frac{1}{16\pi G}(R+\zeta R_{\mu\nu}B^\mu B^\nu)-\frac{1}{4}B_{\mu\nu}B^{\mu\nu}-V(B_\mu B^\nu \pm b^2) \right],
\end{equation}
where $\zeta$ is a real constant that controls the non-minimal coupling between the bumblebee field $B_{\mu}$ and curvature $R_{\mu\nu}$;  $b^2$ is a real positive constant. The bumblebee field strength is defined as
$B_{\mu\nu}= D_\mu B_\nu - D_\nu B_\mu$
where $D_\mu$ is the covariant derivative.
The vanishing potential, $V( B^{\mu} B_\mu \pm b^2) = 0$ will give the determination of the vacuum expectation value of the bumblebee field. In this case, we have $B^{\mu} B_\mu \pm b^2=0$, and the bumblebee field vacuum expected value is $b^\mu$, with $b^\mu b_\mu = \mp b^2$. Note that $b_\mu$ is the coefficient of Lorentz and CTP violation, and the $\pm$ signs in the potential determine that $ b_ \mu $ is timelike or spacelike.

The Schwarzschild-like solutions to the above action were given in \cite{ Bertolami:2005bh, Casana:2017jkc} which reduce to the Schwarzschild black hole when the vacuum expectation value of the bumblebee field is null. The metric of a Kerr-like geometry which satisfies the Einstein equation in Einstein bumblebee theory was given in \cite{Ding:2019mal}, but with that metric the  bumblebee field equation cannot be fulfilled unless it comes the slowly rotating limit \cite{Ding:2020kfr,Maluf:2022knd,Kanzi:2022vhp}. Therefore, the general rotating black hole solution is still an open question for the Einstein bumblebee gravity theory, but the slowly rotating Kerr-like black hole solution which satisfies the equations of motion can be written as \cite{Ding:2020kfr,Maluf:2022knd,Kanzi:2022vhp}
\begin{equation}\label{eq:Kerr-like metric}
ds^2=-\left(1-\frac{2M}{r}\right)dt^2+\frac{(1+L)r}{r-2M}dr^2+r^2d\vartheta^2+r^2 \sin^2\vartheta d\varphi^2-\frac{4Ma\sqrt{1+L}\sin^2\vartheta}{r}dtd\varphi.
\end{equation}
In this solution, $M$ and $a$ are the mass and spin momentum of the black hole, respectively. $L$ is the radial Lorentz violation parameter which is connected with vacuum expectation of the bumblebee field. For vanishing $a$, the above solution reduces to the Schwarzschild-like solution in \cite{Casana:2017jkc}, while for $L=0$, it recovers slowly rotating Kerr black hole in Einstein's general relativity.

\section{Strong gravitational lensing effect}\label{sec:SGL}
In this section, we shall study the gravitational lensing by the slowly rotating Kerr-like black hole and explore the effect of Lorentz violation on the lensing observables in the strong field limit.

\subsection{Light rays in the equatorial plane}\label{sec:III.A}
To explore the gravitational lensing effect, here we shall closely follow the process of \cite{Bozza:2003cp} and consider only light rays in the equatorial plane $(\vartheta=\pi/2)$. Then we rewrite the slowly rotating Kerr-like metric \eqref{eq:Kerr-like metric} projected on the equatorial plane as
\begin{eqnarray}\label{eq:Projected metric}
ds^2=-A(x)\,dt^2+B(x)\,dx^2 +C(x)\,d\varphi^2-D(x)dt\,d\varphi,
\end{eqnarray}
where
\begin{eqnarray}
A(x)&=& 1-\frac{1}{x},\;\;\;\;\;~~~~
B(x)=\frac{(1+L) x}{(x-1)},
\nonumber\\
C(x)&=& x^2,
\;\;\;~~~~~~~~
D(x)=\frac{2a\sqrt{1+L}}{x}.
\end{eqnarray}
where we have rescaled all the quantities $r,a,t$ in the units of $2M$ and set $M=1/2$, and also use $x$ instead of $r$.

We shall use the Hamilton-Jacobi method \cite{Carter:1968rr} to analyze the motions of photon's trajectory. The Lagrangian of photons writes $\mathcal{L}=\frac{1}{2}g_{\mu\nu}\dot{x}^{\mu}\dot{x}^\nu$=0 where the dot represents the derivative with respect to the affine parameter $\lambda$. Then we introduce the Hamilton-Jacobi equation
\begin{equation}
\mathcal{H}=-\frac{\partial S}{\partial \lambda}=\frac{1}{2}g_{\mu\nu}\frac{\partial S}{\partial x^{\mu}}\frac{\partial S}{\partial x^{\nu}}=0,
\label{Lagrangian}
\end{equation}
where $\mathcal{H}$ and $S$ are the canonical Hamiltonian and the Jacobi action. Since the metric \eqref{eq:Projected metric} have two linearly independent killing vector associated with the time translation and rotational invariance, the photon's trajectory is determined by two conserved quantities, namely the energy and angular momentum which are defined as
\begin{eqnarray}
E:=-\frac{\partial S}{\partial t}=-g_{\varphi t}\dot{\varphi}-g_{tt}\dot{t},~~~\mathrm{and}~~~
L_z:=\frac{\partial S}{\partial \varphi}=g_{\varphi\varphi}\dot{\varphi}+g_{\varphi t}\dot{t}~.
\label{momentum}
\end{eqnarray}
We then choose the suitable $\lambda$ such that $E=1$, after replacing $L_z$ by $u$,  we have the following equations of motion for the photons
\begin{eqnarray}
\dot{t} &=& \frac{4C-2 u D}{4AC + D^2},\label{eq:tdot}\\
\dot{\varphi} &=& \frac{2D+4 A u}{4AC + D^2},\\
\dot{x} &=& \pm 2 \sqrt{\frac{C-Du-Au^2}{B(4AC + D^2)}},\label{eq:xdot}
\end{eqnarray}
where we abbreviate the $x$-dependent metric functions to $A,B,C$ and $D$.
The effective potential for the radial motion is defined as
\begin{eqnarray}\label{eq:effpot}
V_{\text{eff}}:=\dot{x}^2 = \frac{4(C-Du-Au^2)}{B(4AC + D^2)},
\end{eqnarray}
which determines the different types of photon orbits. In details, when $V_{\text{eff}}(x=x_0)=0$, a light ray from the source may turn at the radius $x_0$ which is denoted as the closest or minimal approach distance toward the black hole, and then go towards to the observer. Subsequently, we can reduce the impact parameter $u$ as
\begin{equation}\label{eq:u}
u(x_0)=\frac{-D_0+\sqrt{4A_0C_0+D_0^2}}{2A_0}
\end{equation}
where the functions with subscript $0$ are evaluated at $x=x_0$.
Moreover, as known that the deflection angle will diverge when the light approaches the photon sphere with radius $x_0=x_m$, which is determined by
\begin{eqnarray}
V_{\text{eff}}=\frac{dV_{\text{eff}}}{dx}\Big|_{x_0=x_m}=0.
\end{eqnarray}
Moreover, the unstable photon sphere should also satisfy $\frac{d^2 V_{\text{eff}}}{dx^2}\Big|_{x_0=x_m}<0$ \cite{Harko:2009xf}. Therefore, the photon orbit radius $x_m$ is the largest root of the following equation
\begin{eqnarray}\label{eq:ps}
AC'-A'C + u(A'D - AD') = 0,
\end{eqnarray}
where the prime denotes the derivative to $x$.

\begin{figure}[H]
{\centering
\includegraphics[scale=0.5]{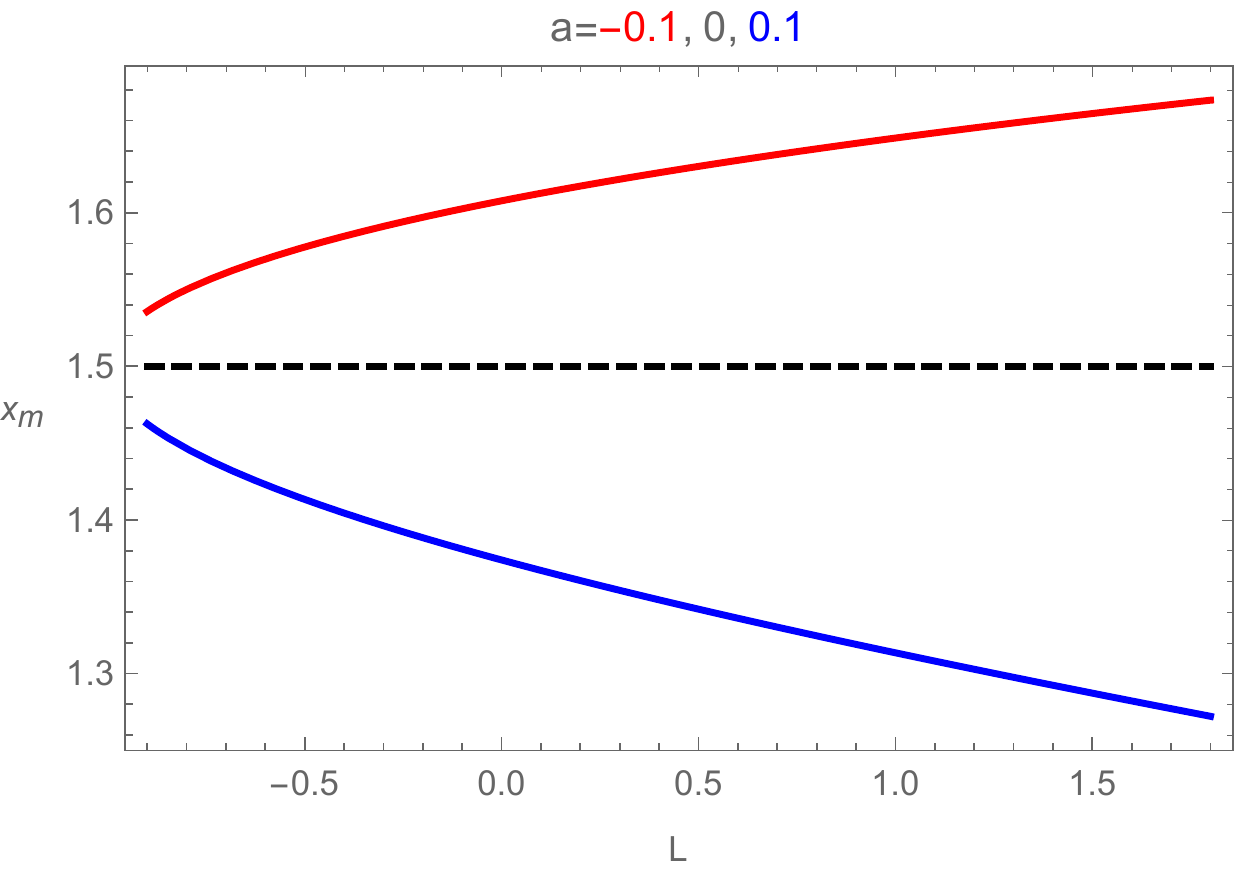}
\caption{The radius of the photon sphere as a function of LV parameters with different spin parameters for the slowly rotating Kerr-like black hole. The dashed line denotes the static case, while solid curves are for direct photons (blue) and retrograde photons (red), respectively.}\label{fig:a-xm}	}	
\end{figure}

In FIG.\ref{fig:a-xm}, we show $x_m$ for the black hole parameters, the LV parameter $L$ and the rotation parameter $a$.  Increasing $L$ decreases the radius of the photon sphere for positive $a$,  but grows the photon sphere for negative $a$; while for static case, $x_m=3/2$ which is the same as in Schwarzschild black hole, independent on $L$. On the other hand, the figure also shows that the photon sphere  becomes smaller as the value of $a$ increases, which reproduces the result for Kerr black hole in \cite{Islam:2021dyk}. This result indicates that for positive $a$, the LV will promote the photons to get closer to the black hole while for the negative $a$, the situation completely changes.  We shall see that this features will be reflected by the lensing observables. Note that here we fix the counterclockwise winding of light rays, so  positive $a$ indicatas that the black hole also rotates in the
the same direction of photon winding and we observe the direct photons, while negative $a$ means that the black hole rotates in clockwise direction and we observe the retrograde photons.

\subsection{Deflection angle}
According to \cite{Bozza:2003cp}, the deflection angle by the rotating black hole \eqref{eq:Projected metric}, at $x_0$ is evaluated by
\begin{eqnarray}\label{bending1}
\alpha(x_0)=I(x_0)-\pi,
\end{eqnarray}
where
\begin{eqnarray}\label{bending2}
I(x_0) = 2 \int_{x_0}^{\infty}\frac{d\varphi}{dx} dx
= 2\int_{x_0}^{\infty}\frac{\sqrt{A_0 B }\left(2Au+ D\right)}{
\sqrt{4AC+D^2}\sqrt{A_0 C-A C_0+u\left(AD_0-A_0D\right)}} dx.
\end{eqnarray}
The above integral usually can not be solved easily. An effective way to handle the integral is to expand the deflection angle in the strong deflection limit near the photon sphere \cite{Bozza:2002zj,Tsukamoto:2016jzh}, which could give an analytical presence of the deflection angle.
Subsequently, to evaluate the integral, we introduce the invariable $z=\frac{A-A_0}{1-A_0}$, and rewrite the integral as
\begin{eqnarray}\label{integral}
I(x_0)=\int_{0}^{1} R(z,x_0)f(z,x_0)dz,
\end{eqnarray}
where
\begin{eqnarray}
R(z,x_0)=\frac{2(1-A_0)}{A'} \frac{\sqrt{B}\left(2A_0Au+A_0D\right)}{\sqrt{CA_0}\sqrt{4AC+D^2}},
\end{eqnarray}
\begin{eqnarray}\label{fz}
f(z,x_0)= \frac{1}{\sqrt{A_0-A\frac{C_0}{C}+\frac{u}{C}\left(AD_0-A_0D\right)}}.
\end{eqnarray}
 The function $R(z,x_0)$ is  regular for all values of $z$ and $x_0$, but  $f(z,x_0)$ diverges as $z \to 0$. Thus, to proceed, we expand the argument of square root in $f(z,x_0)$ to the second order, then the function $f(z,x_0)$ can be approximated as
\begin{eqnarray}
f(z,x_0)\sim f_0(z,x_0)=\frac{1}{\sqrt{\mathfrak{m}(x_0) z + \mathfrak{n}(x_0)z^2}},
\end{eqnarray}
where $\mathfrak{m}(x_0)$ and $\mathfrak{n}(x_0)$ are the coefficient of Taylor expansion.

Then the integral will give the strong field limit of the deflection angle as \cite{Bozza:2002zj,Bozza:2003cp}
\begin{eqnarray}\label{eq:alpha-def}
\alpha(u)=-\bar{a} \log\Big(\frac{u}{u_m}-1\Big)+ \bar{b} + \mathcal{O}\left(u-u_m\right),
\end{eqnarray}
where $u$ is the impact parameter defined in \eqref{eq:u} and
$u_m=u(x_m)$.
The strong deflection  coefficients $\bar{a}$ and $\bar{b}$ are
\begin{eqnarray}\label{eq:abarbbar}
\bar{a} = \frac{R(0,x_m)}{2\sqrt{{\mathfrak{n}}_m}}, ~~~ \textrm{and}~~~ \bar{b} = -\pi +b_D+b_R + \bar{a} \log\frac{\bar{c} x_m^2 }{u_m^2}
\end{eqnarray}
where
\begin{eqnarray}\label{eq:abar-coeffi}
b_D=2\bar{a}\log\frac{2(1-A_m)}{A'_mx_m},~~~~~
b_R= \int_{0}^{1} [R(z,x_m)f(z,x_m)-R(0,x_m)f_0(z,x_m)]dz,
\end{eqnarray}
respectively, and $\bar{c}$ is defined by the coefficient in Taylor expansion
\begin{equation}\label{eq:c}
u-u_m=\bar{c}(x_0-x_m)^2.
\end{equation}
Note that the functions with the subscript $m$ are evaluated at $x=x_m$.

With the expression \eqref{eq:alpha-def}-\eqref{eq:c} in hands, we are ready to study the deflection angle of the strong lensing by the slowly rotating Kerr-like black hole in bumblebee gravity to distinguish it from the Kerr black hole and may probe the
value of the LV constant $L$ by using strong field gravitational lensing.

\begin{figure}[H]
{\centering
\includegraphics[scale=0.45]{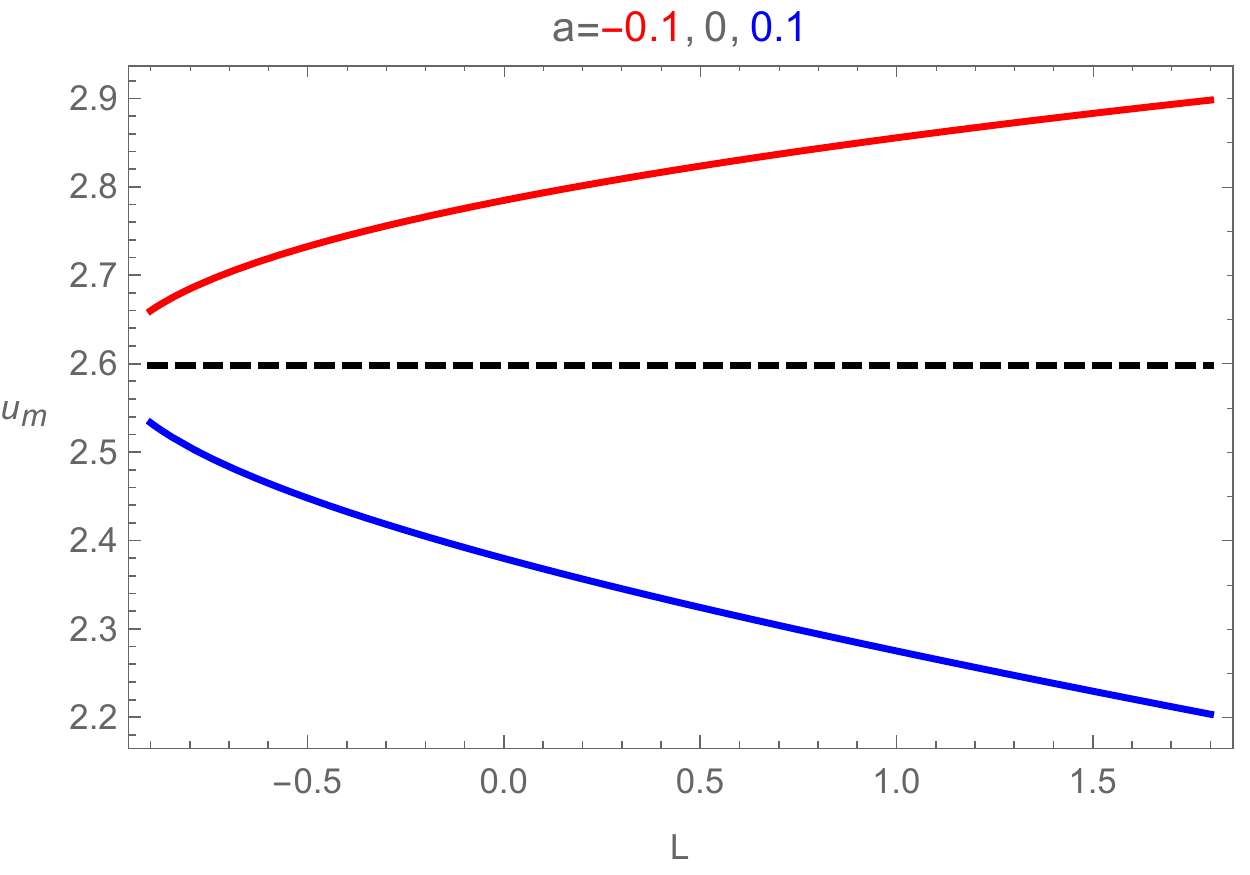}\hspace{0.2cm}
\includegraphics[scale=0.45]{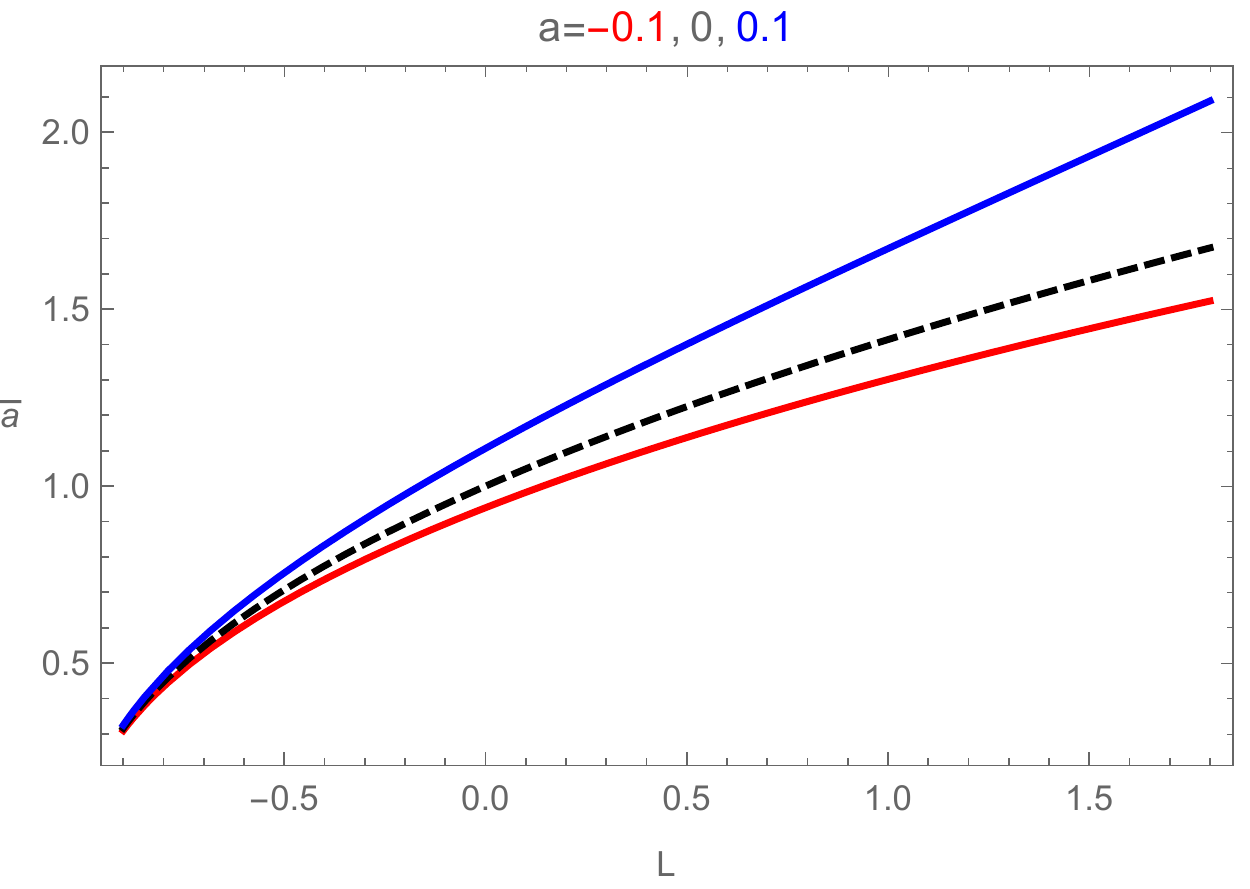}\hspace{0.2cm}
\includegraphics[scale=0.45]{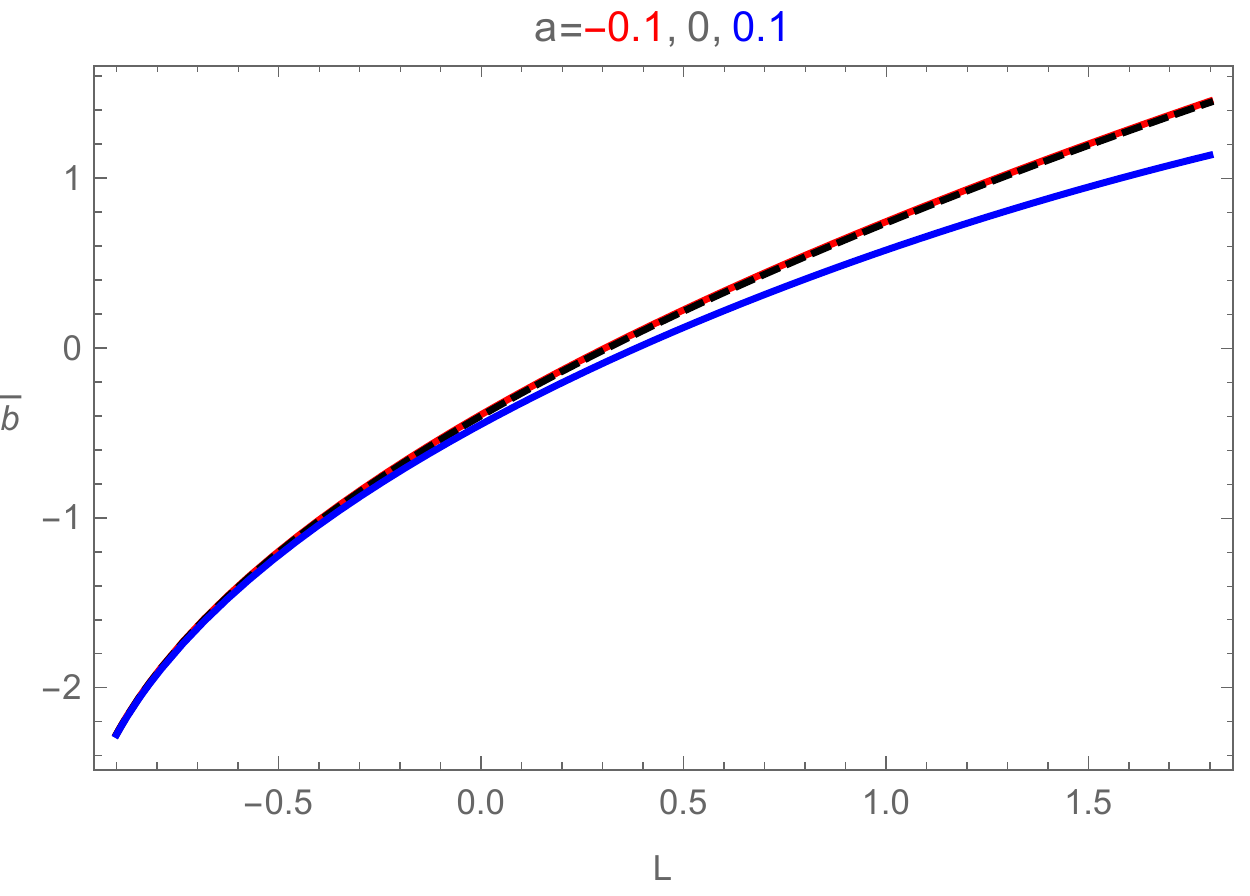}
\caption{The behavior of the critical impact parameter $u_m$ (left panel)  and the lensing coefficient  $\bar{a}$ (middle panel) and $\bar{b}$ (right panel) as functions of the LV parameter with different spin parameters for the slowly rotating Kerr-like black hole. In each panel, the dashed line denotes the static case, while solid curves are for direct photons (blue) and retrograde photons (red), respectively.}\label{fig:a-um}	}	
\end{figure}

In FIG.\ref{fig:a-um}, we study the strong field deflection coefficients, $u_m$, $\bar{a}$, and $\bar{b}$, plotted against the LV parameter $L$.  It shows that $u_m$, $\bar{a}$ and $\bar{b}$ for Kerr-like black hole could larger or smaller than those for Kerr black hole, depending on the sign of the spin parameter.
In details, the dependence of $u_m$ on $L$ and $a$ are qualitatively same as that of $x_m$.  Note that if the impact parameter $u < u_m$, the photon rays fall into the black hole and for $u>u_m$ the photon rays scattered by black hole. Both $\bar{a}$ and $\bar{b}$   increase as $L$ increases. Moreover,  the difference of $\bar{a}$ and $\bar{b}$  between the negative $a$ and the static case is smaller than that between the positive $a$ and the static case,  in particular, the value of $\bar{b}$ for negative $a$ is almost the same as that for static case.

\begin{figure}[H]
{\centering
\includegraphics[scale=0.5]{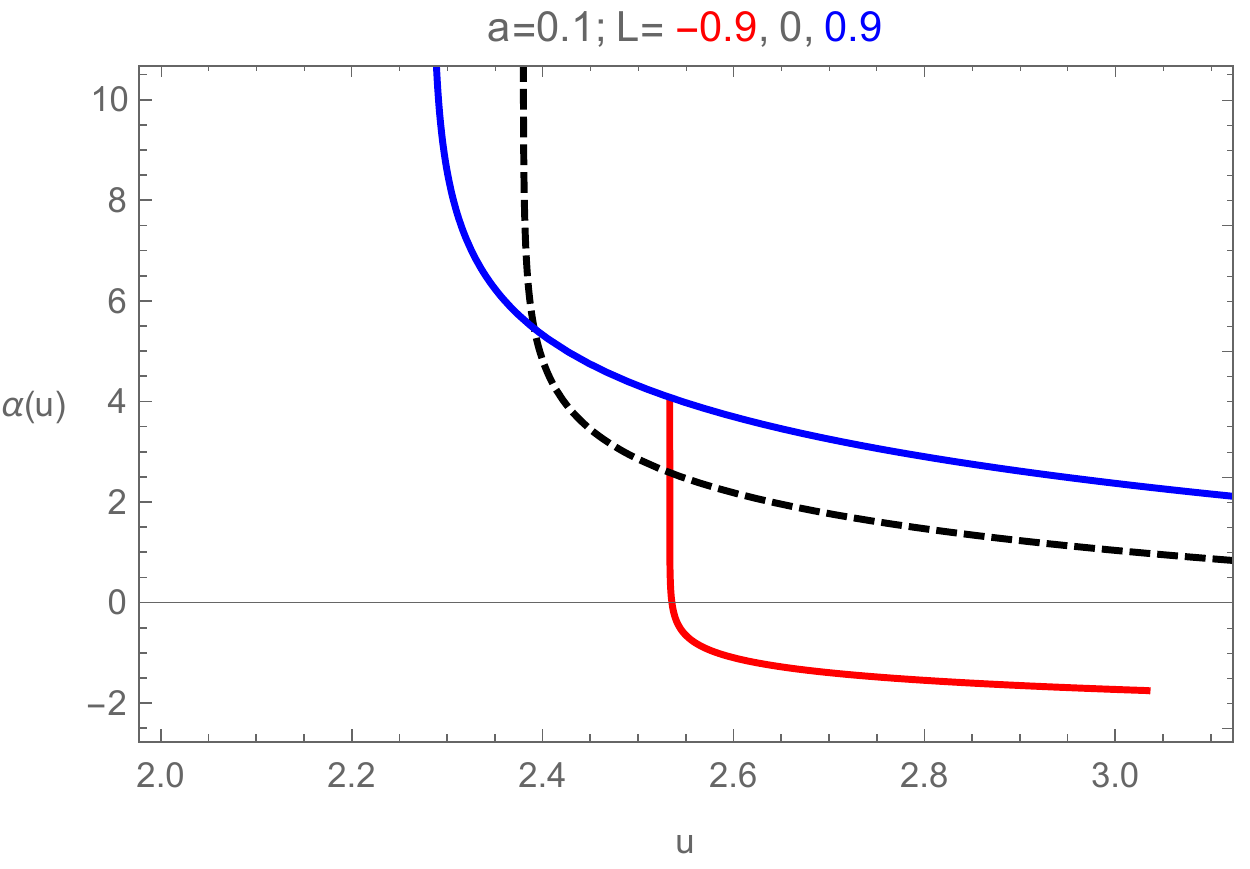}\hspace{1cm}
\includegraphics[scale=0.5]{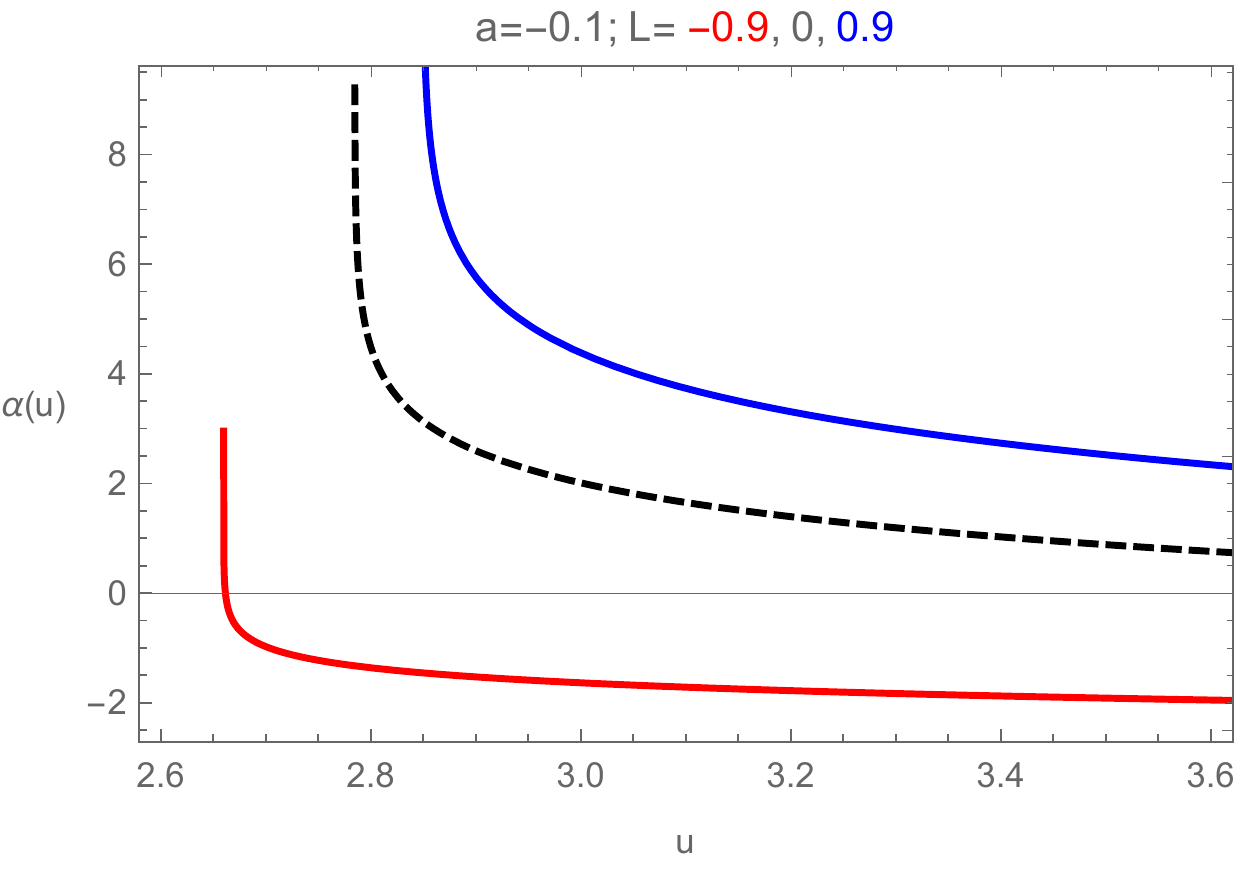}
\caption{The deflection angle as a function of impact parameter with different LV parameters. The left panel is for direct photons  while the right panel is for retrograde photons.}\label{fig:u-alpha}	}	
\end{figure}

The imprint of the parameters, $L$ and $a$ on the coefficients can be partly inferred from the deflection angle present in FIG.\ref{fig:u-alpha}. The deflection angle diverges at a smaller $u_m$ for smaller $L$  when $a$ is positive (left plot), but for larger $L$ when $a$ is negative (right plot). In all cases, the deflection angle is monotonically decreasing with increasing the impact parameter $u$, and the strong lensing is valid only when $u$ is close to $u_m$, especially for smaller $L$. Moreover, the LV parameter can promote or suppress the deflection angle comparing to the Kerr case ($L=0$), in particular, the deflection angle for negative $a$ monotonously depends on $L$ but there exists a intersection for positive $a$, which could attribute to the different behavior of $\bar{a}$ and $\bar{b}$ for different sign of $a$ depicted in  FIG.\ref{fig:a-um}.

\subsection{Observables in the relativistic images by M87*}

\begin{figure}[H]
{\centering
\includegraphics[scale=0.45]{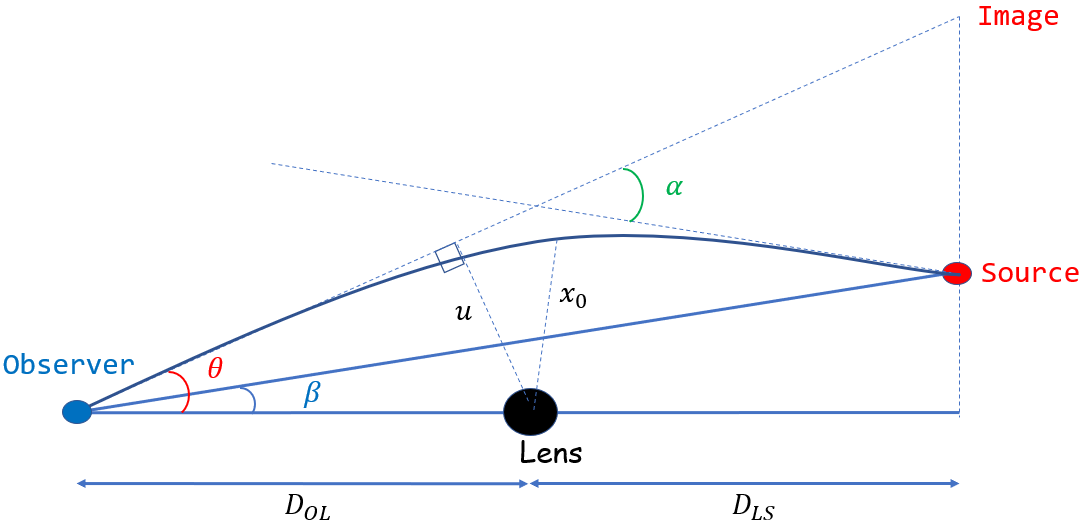}
\caption{The geometrical configuration of gravitational lensing. }\label{fig:lensingDiagram}	}	
\end{figure}

Considering that a light ray emitted by a source is scattered by a black hole or lens, the connection
among the black hole (lens), observer and the light source is governed by the lens equation.  One usually assume that the source and observer are far from the black hole and they are perfectly aligned, but it is better to start from the equation for small lensing angle reads \cite{Bozza:2001xd}
\begin{equation}\label{eq:lenseq}
\beta=\theta-\frac{D_{LS}}{D_{OS}}\Delta\alpha_{n},
\end{equation}
where  $\Delta\alpha_{n}=\alpha-2n\pi$ is the offset of deflection angle looping over $2n\pi$ and $n$ is an integer. Here, $\beta$ is the angular separations between the source and the black hole, and $\theta$ is the angular separations between the image and the black hole. $D_{OL}$ is the distance between the observer and the lens, and $D_{OS}$ is the distance between the observer and the source.
A cartoon of geometrical configuration about gravitational lensing is shown in FIG. \ref{fig:lensingDiagram}.

\subsubsection{Review on the observables in lensing }

Using \eqref{eq:alpha-def} and \eqref{eq:lenseq}, the position of the $n$-th relativistic image can be approximated as \cite{Bozza:2002zj}
\begin{equation}
\theta_n=\theta^0_n+\frac{u_{m}e_n(\beta-\theta^0_n)D_{OS}}{\bar{a}D_{LS}D_{OL}},
\end{equation}
where
\begin{equation}
e_n=\text{exp}\left({\frac{\bar{b}-2n\pi}{\bar{a}}}\right)
\end{equation}
and $\theta^0_n$ are the image positions corresponding to
$\alpha=2n\pi$. Since the gravitational lensing conserves surface brightness, the magnification is the quotient of the solid angles subtended by the $n$-th image, and the source \cite{Bozza:2002zj,Virbhadra:1999nm,Virbhadra:2008ws}. Thus, the magnification of $n$-th relativistic image is evaluated  as \cite{Bozza:2002zj}
\begin{equation}\label{mag}
\mu_n=\left(\frac{\beta}{\theta} \;  \;\frac{d\beta}{d\theta} \right)^{-1}\Bigg|_{\theta_n ^0} = \frac{u^2_me_n(1+e_n)D_{OS}}{\bar{a}\beta D_{LS}D^2_{OL}}.
\end{equation}
It is obvious that the magnifications could decrease exponentially with $n$ and the first relativistic image is the brightest one. Moreover, the magnifications are proportional to 1/$D_{OL}^2$ which is very small, and thus the relativistic images are very faint, but if the values of $\beta$ are close to zero, i.e. nearly perfect alignment, the images can be bright enough.

Given that $\theta_{\infty}$ represents the asymptotic position of a set of images in the limit $n\rightarrow \infty$, we can treat the outermost image $\theta_1$ as a single image and all the remaining ones packed together as $\theta_{\infty}$. Then, combing the deflection angle \eqref{eq:alpha-def} and lens equation \eqref{eq:lenseq}, we can evaluate the three lensing observables, i.e. the angular position of the asymptotic relativistic images ($\theta_{\infty}$), angular separation between the outermost and asymptotic relativistic images ($s$) and relative magnification of the outermost relativistic image with other relativistic images, ($r_{\text{mag}}$) as \cite{Bozza:2002zj,Islam:2021ful}
\begin{align}
\theta_{\infty}& = \frac{u_{m}} {D_{OL}},\\
s& = \theta_1-\theta_{\infty}=\theta_\infty ~\text{exp}\left({\frac{\bar{b}}{\bar{a}}-\frac{2\pi}{\bar{a}}}\right),\label{eq:s}\\
r_{\text{mag}}& =\frac{\mu_1}{\sum{_{n=2}^\infty}\mu_n }\simeq \frac{5\pi}{\bar{a}~\text{log}(10)}\label{eq:mag1}.
\end{align}

Moreover, if one can distinguish the time signals of the first image and other packed images, then one can consider another important observable in strong field lensing, i.e., the time delay. The deflection angle for the slowly rotating Kerr-like black holes could be more than $2\pi$, and multiple images of the source could be formed. So, the time travelled by the light paths corresponding to the different images is not the same,  therefore, it would have a time difference between the two images. When the $p-$th and $q-$th images are on the same side of the lens, the time delay between them could be approximated as  \cite{Bozza:2003cp}
\begin{eqnarray}\label{eq:timedelay}
\Delta T_{p,q} \approx 2\pi(p-q) \frac{\widetilde{R}(0,x_m)}{2\bar{a}\sqrt{{\mathfrak{n}}_m}} + 2\sqrt{\frac{A_m u_m}{B_m}}\left[ e^{(\bar{b}-2q\pi\pm \beta)/2\bar{a}}-e^{(\bar{b}-2p\pi\pm \beta)/2\bar{a}}\right]
\end{eqnarray}
where
\begin{eqnarray}
\widetilde{R}(z,x_m) &=&\frac{2(1-A_0)\sqrt{BA_0}[2C-u D]}{A'\sqrt{C(D^2+ 4AC)}}\left(1- \frac{1}{\sqrt{A_0}f(z,x_0)}\right).
\end{eqnarray}
In \eqref{eq:timedelay}, the value of the time delay mainly stems from the contribution of the first term because the second term could be  negligible.  Thus, the time delay \eqref{eq:timedelay} becomes \cite{Bozza:2003cp}
\begin{eqnarray}\label{eq:td2}
\Delta T_{p,q} \approx 2\pi(p-q)\frac{\widetilde{R}(0,x_m)}{2\bar{a}\sqrt{{\mathfrak{n}}_m}}=2\pi(p-q)\frac{\tilde{a}}{\bar{a}}
\end{eqnarray}
where
\begin{eqnarray}
\tilde{a} = \frac{\widetilde{R}(0,x_m)}{2\sqrt{{\mathfrak{n}}_m}}.
\end{eqnarray}
Since as pointed out in \cite{Bozza:2003cp} that the time delay for direct photons ($a>0$) is different from the prograde photons ($a<0$), we could consider the case when the two images are at opposite sides of the lens, which is distinguishable from the above case. In this case, the time delay is given as  \cite{Bozza:2003cp}
\begin{eqnarray}\label{eq:timedelayo}
\Delta \widetilde{T}_{p,q} \approx \frac{\tilde{a}(a)}{\bar{a}(a)}[2\pi p +\beta-\bar{b}(a)]+\tilde{b}(a)-\frac{\tilde{a}(-a)}{\bar{a}(-a)}[2\pi q +\beta-\bar{b}(-a)]-\tilde{b}(-a),
\end{eqnarray}
where
\begin{eqnarray}
\tilde{b} = -\pi +\tilde{b}_D(x_m)+ \tilde{b}_R(x_m) + \tilde{a}~\text{log}\left( \frac{\bar{c} x_m^2}{u_m}\right),
\end{eqnarray}
with
\begin{eqnarray}
\tilde{b}_D=2\tilde{a}\log\frac{2(1-A_m)}{A'_mx_m},~~~~~
\tilde{b}_R(x_m) = \int_{0}^{1} [\widetilde{R}(z,x_m)f(z,x_m)-\widetilde{R}(0,x_m)f_0(z,x_m)]dz.
\end{eqnarray}

Thus, we see that in theoretical aspect, we can predict the above observables of the strong lensing by the slowly rotating Kerr-like black hole \eqref{eq:Kerr-like metric} from various lensing coefficients. Inversely, in experimental side, if one can observe
 $s$, $r_{\text{mag}}$,  $\theta_{\infty}$ and the time delays, then this would be helpful to identify the nature of the rotating Kerr-like black holes or lens.
 Next, we shall investigate the gravitational lensing by treating the slowly Kerr-like black hole as the supermassive M87* black hole, and do the comparison with the Kerr case in GR.

\subsubsection{Gravitational lensing for suppermassive black hole in M87*}
\begin{figure}[H]
{\centering
\includegraphics[scale=0.45]{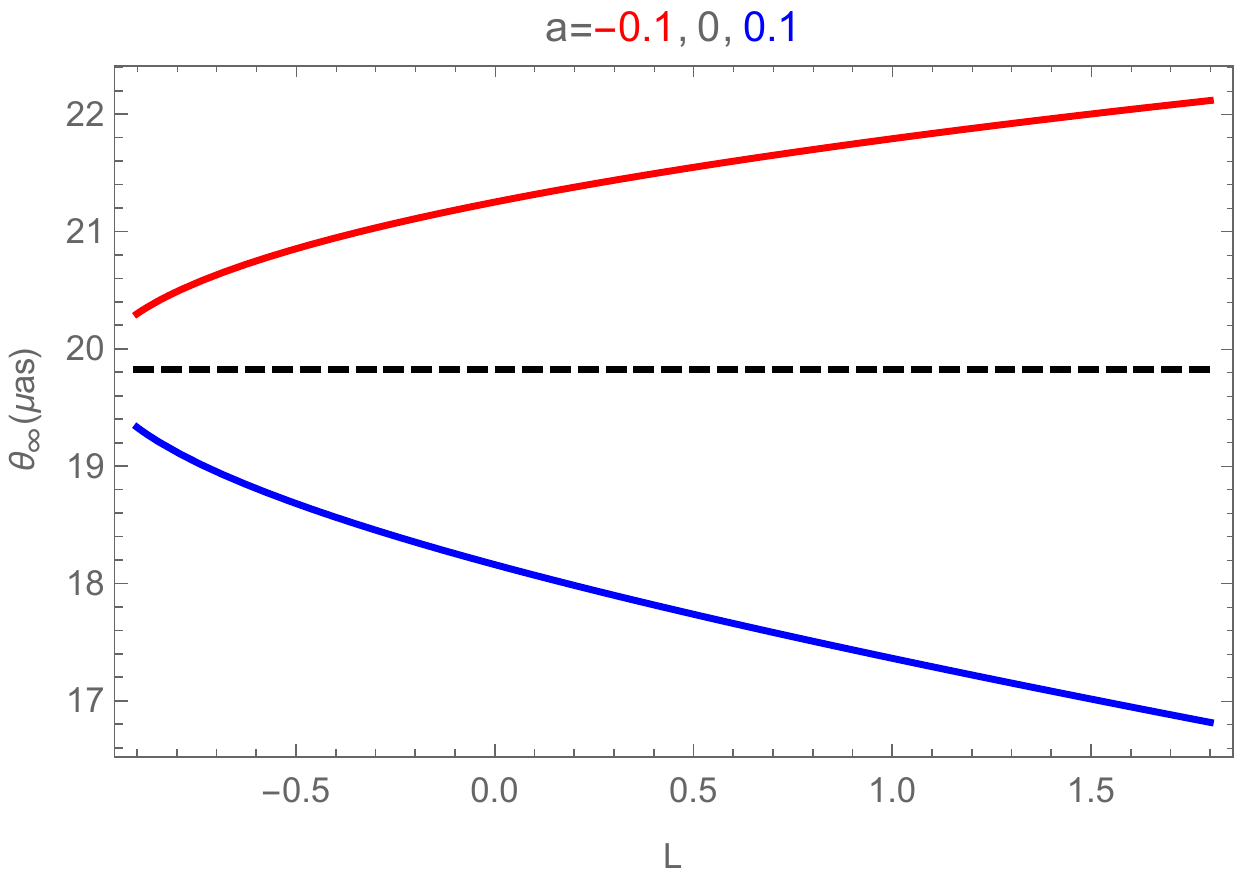}\hspace{0.2cm}
\includegraphics[scale=0.45]{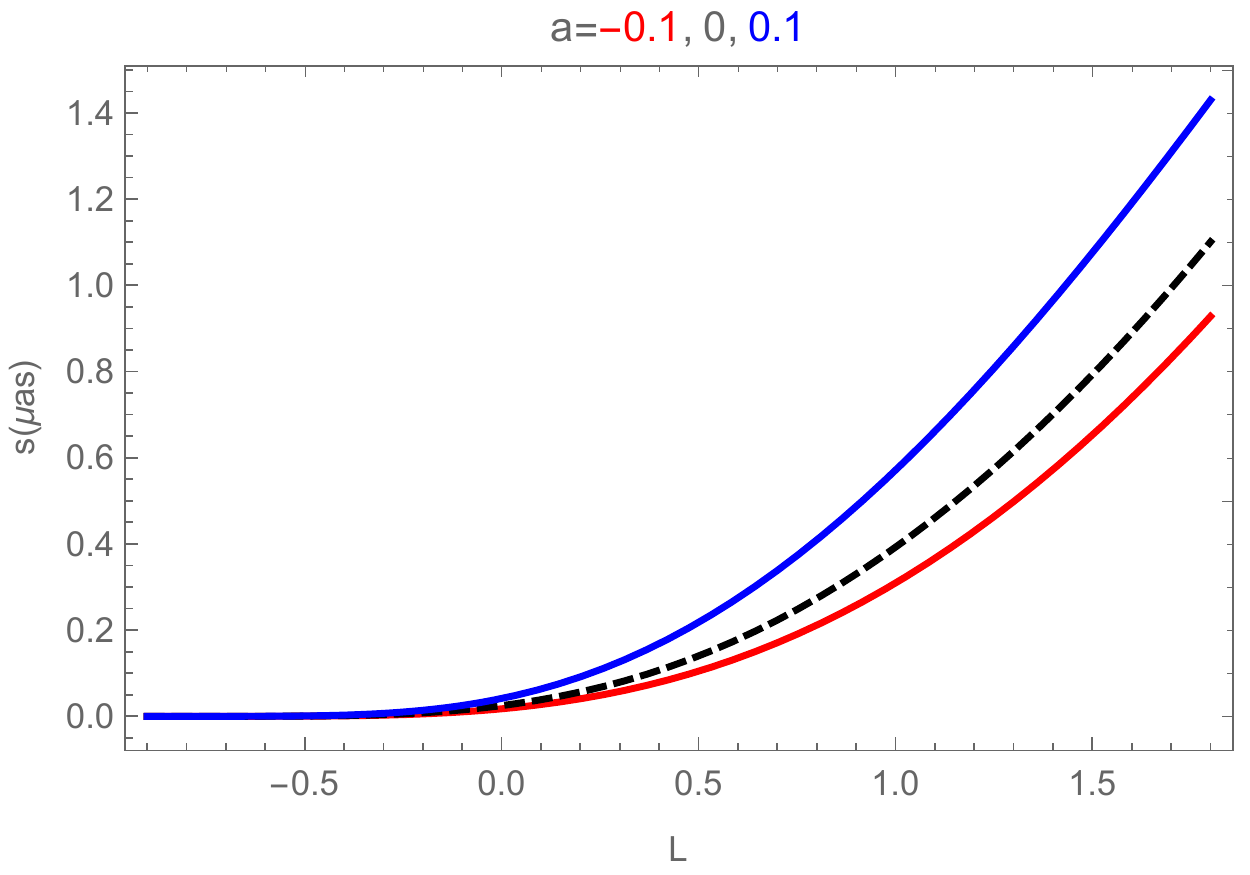}\hspace{0.2cm}
\includegraphics[scale=0.45]{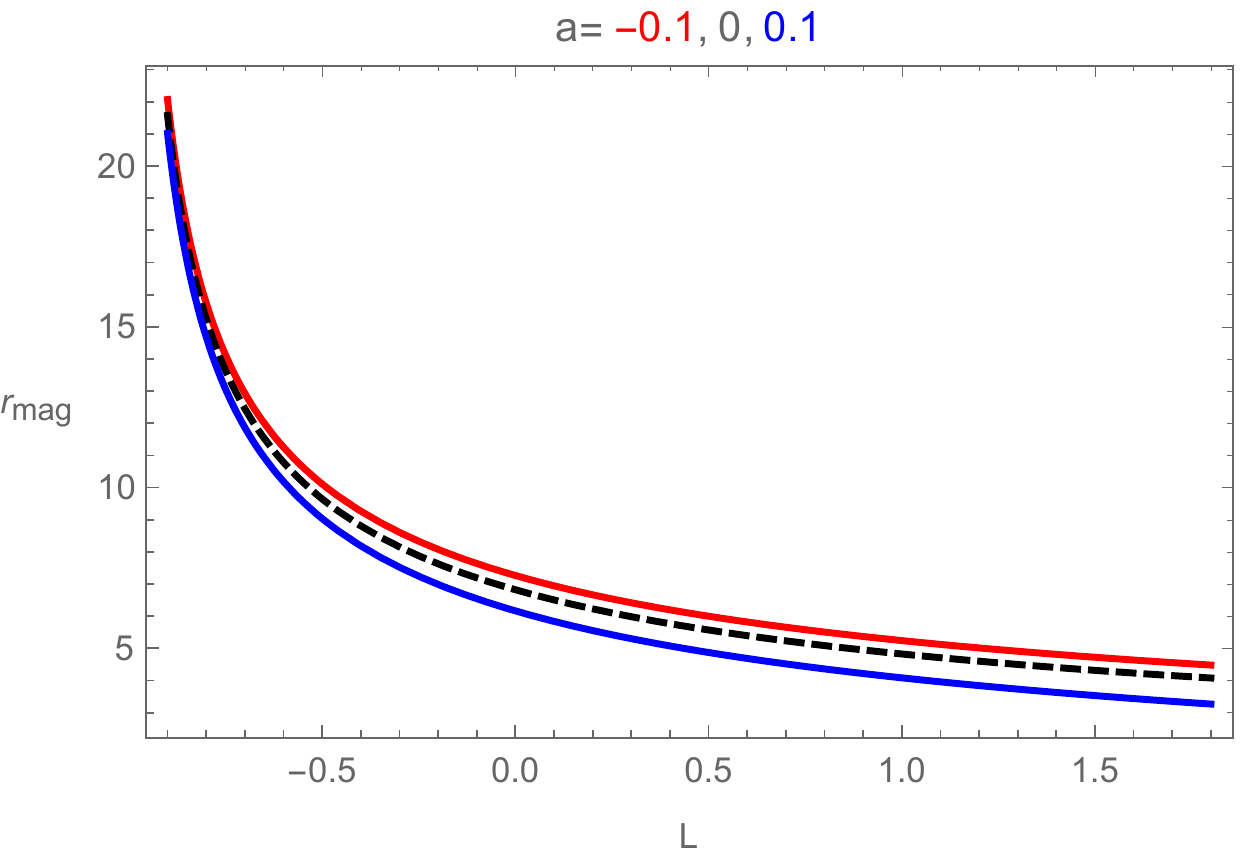}
\caption{The behavior of lensing observables, $\theta_\infty$ (left panel), $s$ (middle panel) and $r_{\text{mag}}$ (right panel) chaning with LV parameter in strong gravitational lensing by considering the rotating Kerr-like black hole as the M87* black hole. In each panel, the dashed line denotes the static case, while solid curves are for direct photons (blue) and retrograde photons (red), respectively.}\label{fig:observables}	}	
\end{figure}

Presupposing the supermassive black hole M87* as the lens by the Kerr-like black hole, with mass $M=6.5\times 10^{9}M_{\odot}$ and $D_{OL}=16.8 Mpc$,  we shall numerically evaluate the lensing observables $\theta_\infty$, $s$ and $r_{\text{mag}}$ and the time delays of the  rotating Kerr-like black hole, depicted in FIG. \ref{fig:observables}, TABLE \ref{table01} and TABLE \ref{table02}, respectively. Then we focus on the effect of the LV parameter and the deviation of the observables in Einstein-bumblebee gravity than those in GR ($L=0$).

In FIG. \ref{fig:observables}, it shows that the observable, $\theta_\infty$, for Kerr-like black hole in Einstein-bumblebee gravity is significantly larger/smaller than the Kerr case when rotation parameter $a$ is negative/positive for increasing values of LV parameter $L$, the same as the behavior of $u_m$ as we expect from its definition. The deviation from the Kerr case of the observable $s$, is slight for negative $L$, and become significant for larger $L$, though the deviation value depends on the sign of $a$. In addition, \eqref{eq:mag1} means that the observable $r_{mag}$ is inversely proportional to $\bar{a}$, and so the behavior of $r_{mag}$ in the right plot is in contrary to that of $\bar{a}$, i.e, comparing with Kerr case, the image for negative $L$ is more brighter while it is more fainter for positive $L$.

These studies indicate the deviation of the characterized observables in Kerr-like black hole from that in Kerr black hoke, but the difference is very small in the order of $\mu as$, which is beyond the current EHT observation. We still expect that the developing technics could improve the precision of the observation and then help us to distinguish the bumblebee theory from GR.

\begin{table}[!htp]
{\centering
\begin{tabular}{|c|c|c|c|c|c|c|c|}
  \hline
\diagbox{a}{$\frac{\Delta T_{2,1}}{\mathrm{hrs}}$}{L} & -0.9 & -0.6 & -0.3 & 0 & 0.3 & 0.6 & 0.9\\
  \hline
  -0.1 & 296.650 &303.240 &307.354 & 310.571 & 313.283 & 315.659 & 317.794\\
  \hline
  -0.05 & 293.245& 296.650 & 298.808 & 300.514 & 301.962& 303.240 & 304.394 \\
  \hline
  0 & 289.760 & 289.760 & 289.760 & 289.760 & 289.760 & 289.760& 289.760 \\
  \hline
  0.05 & 286.189 & 282.525& 280.107 & 278.141 & 276.430& 274.889& 273.471\\
  \hline
  0.1 & 282.525 & 274.889 & 269.710 & 265.407 & 261.588& 258.082 & 254.794 \\
  \hline

\end{tabular}
\caption{The time delay of the first image from that of the second image, $\Delta T_{2,1}$, in strong gravitational lensing by considering the rotating Kerr-like black hole as the M87* black hole.
\label{table01} }}
\end{table}
\begin{table}[!htp]
{\centering
\begin{tabular}{|c|c|c|c|c|c|c|c|}
       \hline
\diagbox{a}{$\frac{\Delta \widetilde{T}_{1,1}}{\mathrm{hrs}} $}{L} & -0.9 & -0.6 & -0.3 & 0 & 0.3 & 0.6 & 0.9\\
  \hline
        -0.1& 20.017 & 37.842 & 48.261 & 56.014& 62.279 & 67.570 & 72.166 \\
\hline
      -0.05 &9.999 & 18.847 & 23.957 &27.707 & 30.688 & 33.159 & 35.258 \\
      \hline
       0 & 0 & 0 & 0 & 0 & 0 & 0 & 0 \\
       \hline
       0.05&-9.999  & -18.847 & -23.957 &-27.707 & -30.688 & -33.159 & -35.258 \\
       \hline
       0.1 & -20.017 & -37.842 & -48.261 & -56.014& -62.279  & -67.570 & -72.166 \\
       \hline
\end{tabular}
\caption{The time delay $\Delta \widetilde{T}_{1,1}$ between first order prograde and retrograde images  in strong gravitational lensing by considering the rotating Kerr-like black hole as the M87* black hole.
\label{table02} }}
\end{table}

The time delay of the first image from that of the second image, $\Delta T_{2,1}$ for the slowly rotating Kerr-like black hole is shown in TABLE \ref{table01}. We can extract the properties as follows. (i) For the static case,   $\Delta T_{2,1}==289.760 hrs$ independent of the LV parameter. (ii)For $a<0$, $\Delta T_{2,1}$ is longer for larger $L$ while for $a>0$, $\Delta T_{2,1}$ is shorter for larger $L$. Those properties is reasonable because as pointed as in \cite{Bozza:2003cp}, the main contribution of $\Delta T_{2,1}$ could come from the first term of \eqref{eq:timedelay} which is approximately as $\Delta T_{2,1}\sim 2\pi u_m$. So the dependences of $\Delta T_{2,1}$  on the parameters $L$ and $a$ are similar as that of $u_m$ as shown in FIG.\ref{fig:a-um}.
The time delay $\Delta \widetilde{T}_{1,1}$ between prograde and retrograde images  of the same order is evaluated in
TABLE \ref{table02}. It shows that (i)this kind of time delay has symmetry respect to $a=0$ at which it vanishes , and its magnitude increases when the spin is faster, as we expect. (ii) as $L$ increases, the time delay $\Delta \widetilde{T}_{1,1}$ between prograde and retrograde images of the same order is longer.

To conclude,  comparing to the Kerr case, both $\Delta T_{2,1}$ and $\Delta \widetilde{T}_{1,1}$ could be shorter/longer many hours, which is possible in the current astrophysical observation if one can separate the related two images. Moreover, our theoretical analysis shows that deviation of $\Delta \widetilde{T}_{1,1}$ is sharper than that of $\Delta T_{2,1}$ with the same parameters. On the other hand, as we aforementioned that $\Delta \widetilde{T}_{1,1}$ describe the time delay  between two images staying on opposite sides of lens, which could more easier to be separated than those on the same side. Therefore, we could have more confidence on the measure of $\Delta \widetilde{T}_{1,1}$ to distinguish Einstein bumblebee theory from GR.

\section{Shadow observables and the constraint from M87*}\label{sec:shadow}
In this section, we will  investigate the observables of the slowly rotating Kerr-like black hole shadows which characterize the shape and distortion of the black hole shadow.  Moreover, we will consider the slowly rotating Kerr-like black hole as the black hole M87*, and use the EHT observation to constrain the Lorentz violation parameter. This study could provide us more detailed property of the Einstein bumblebee
theory, and a possible way to evaluate the black hole parameter or to distinguish the Einstein bumblebee gravity from GR. It is noted that some related study on the black hole shadow for the Kerr-like metric which does not satisfy the bumblebee field equation have been done in \cite{Ding:2019mal,Jha:2021bue}, though it may be not proper since the metric is not a solution of the full theory.

\subsection{Null geodesic and photon region}
Similar as in section \ref{sec:III.A}, with
 two conserved quantities \eqref{momentum} and
using the Hamilton-Jacobi method \cite{Carter:1968rr}, we obtain four first-order differential equations for the geodesic motions of the black hole \eqref{eq:Kerr-like metric},
\begin{eqnarray}
&&\dot{t}=1+\frac{2M(r^2-a\sqrt{1+L}\xi)}{r^2(r-2M)},~~~~\dot{r}=\frac{\sqrt{R(r)}}{r^2},\label{eq-motion3}\\
&&\dot{\varphi}=\frac{2Ma\sqrt{1+L}+\xi(r-2M)\csc^2\vartheta}{r^2(r-2M)}, ~~~~
\dot{\vartheta}=\frac{\sqrt{\Theta(\vartheta)}}{r^2} ,
\label{eq-motion4}
\end{eqnarray}
where
\begin{eqnarray}
\Theta(\vartheta)=\eta+\xi^2(1-\csc^2\vartheta),~~~~~~R(r)=\frac{r^4-r(r-2Mr)(\eta+\xi^2)}{1+L}
-\frac{4Mra\xi}{\sqrt{1+L}},
\end{eqnarray}
$\xi=L_z/E$ and $\eta=(K-L_z^2)/E^2$ with $E$ and $L_z$ defined in \eqref{momentum} and  $K$ the Carter constant.

The spherical orbits require $\dot{r}=0$ and $\ddot{r}=0$, which means that $R(r)|_{r=r_p}=0$ and $R'(r)|_{r=r_p}=0$. Subsequently, the constants of motion $\xi$ and $\eta$ are given as
\begin{equation}
\xi=\frac{r_p^2(3M-r_p)}{2Ma\sqrt{1+L}},~~~~
\eta =\frac{r_p^2(12M^2a^2(1+L)-r_p^2(r_p-3M)^2)}{4M^2a^2(1+L)}, \label{eq-ke}
\end{equation}
Substituting the above expression into $\dot{\theta}$ of \eqref{eq-motion3}, the non-negativity of $\Theta$ give us the condition for the photon region
\begin{equation}\label{eq-photonregion}
3r_p^2-\frac{r_p^4(r_p-3M)^2\csc^2\vartheta}{4M^2a^2(1+L)}\geq0.
\end{equation}
For each point in this region, there is a null geodesic staying on the sphere $r=r_p$, along which $\theta$ can oscillate between the extremal values determined by the equality in \eqref{eq-photonregion} and $\varphi$ is governed by \eqref{eq-motion4}.
Moreover, with respect to radial perturbations, the spherical null geodesic at $r=r_p$ is  unstable when $R''(r)|_{r=r_p}>0$ and stable when $R''(r)|_{r=r_p}<0$.

 \subsection{\label{sec:level5B2} Shadow cast and the infalling spherical accretion}
As we discussed, the photon radius could determine the parameters $\xi$ and $\eta$, so the photon region will give us the boundary of the black hole shadow. For observers at infinity distance, according to \cite{Cunningham:cgh}, we can use the celestial coordinate to describe the shadow shape
\begin{eqnarray}
X=\lim_{r_o\to\infty}\left(-r_o^2\sin\vartheta_o\frac{d\varphi}{dr}\Big|_{(r=r_o,\vartheta=\vartheta_o)}\right)~~~\mathrm{and} ~~~~~ Y=\pm\lim_{r_o\to\infty}\left(r_o^2\frac{d\vartheta}{dr}\Big|_{(r=r_o,\vartheta=\vartheta_o)}\right)
\end{eqnarray}
where $(r_o,\vartheta_o)$ is the observer's position in Boyer-Lindquist coordinate. The observer at infinity distance means $r_o\to\infty$ and the inclination angle $\vartheta_o=0     (\mathrm{or}~~\pi)$ corresponds to the observer in north (south) direction while  $\vartheta_o= \pi/2$ corresponds to the observer at equatorial plane of the black hole. Here
the coordinate $X$ is the apparent perpendicular distance of the image as seen from the axis and $Y$ is the apparent perpendicular distance of the image from its projection on the equatorial plane.
In our case, we can further calculate the celestial coordinates as
\begin{eqnarray}\label{eq:X-Y}
X(r_p)&=&-\sqrt{1+L}\xi(r_p)\csc\vartheta_o\nonumber\\
Y(r_p)&=&\pm\sqrt{1+L}\sqrt{\eta(r_p)+\xi(r_p)^2(1-\csc^2\vartheta_o)}
\end{eqnarray}
where the range of $r_p$ is determined by \eqref{eq-photonregion}.
\begin{figure}[H]
{\centering
\includegraphics[scale=0.4]{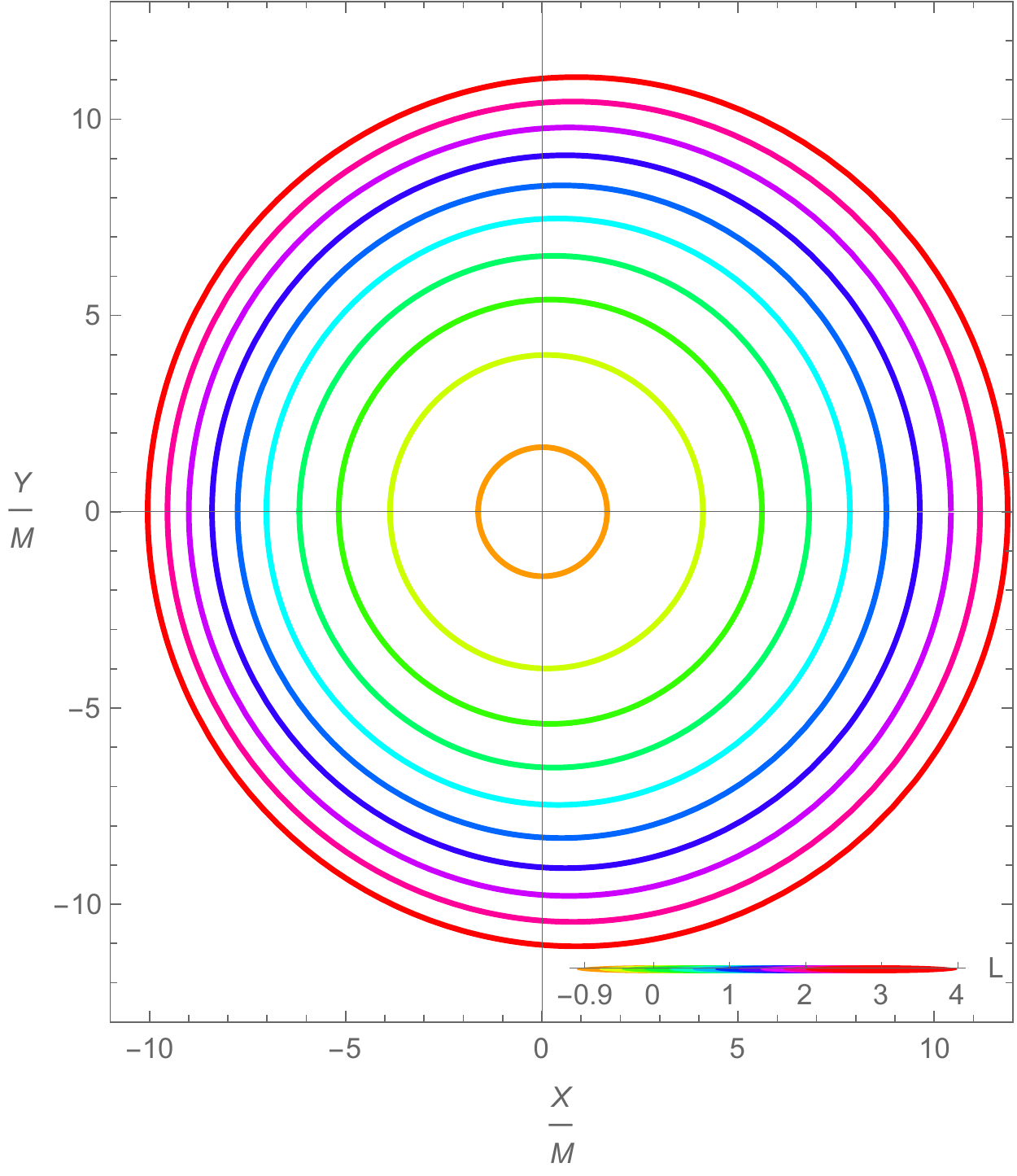}
\caption{Shadow boundary for ten samples of $L$ with equal interval. The inner-most curve is for $L=-0.9$ while the outer-most is for $L=4$. Here we fix $a=0.1$ and $\vartheta_o=\pi/2$.}\label{fig:shadow}	}	
\end{figure}
For the non-rotating case with $a=0$, according to \cite{Perlick:2021aok}, the radius of photon sphere can be solved via $\frac{d(r^2/(1-2M/r))}{dr}=0$  to be $r_{p}=3M$ and the shadow radius is
\begin{equation}
R_{sh}=\sqrt{\frac{r_p^2}{1-2M/r_p}}\big\mid_{r_p=3M}=3\sqrt{3}M
\end{equation}
which are the same as those in Schwarzschild black hole. For the slowly rotating Kerr-like case with small spin, the shadow shape will be deviated from the perfect circle because of the dragging effect. Here we shall focus on the effect of the LV parameter on the black hole shadow, so we fix $a=0.1$ \footnote{From the expressions \eqref{eq-ke} and \eqref{eq:X-Y}, the shadow boundary for different sign of $a$ is symmetric via the vertical axis, so here we will only fix the positive case. Moreover, we shall denote the dimensionless quantity $a/M$ as $a$ for simplicity. } and $\vartheta=\pi/2$.  We show the boundary of shadow for different LV parameters in FIG. \ref{fig:shadow}, which shows that for slowly rotation, the black hole shape is close to circle as expected and its size increases with $L$.

In reality, actual image of the black hole is not seen as the apparent boundary in the universe and within static accretion disk model, but in the more realistic scenario, accretion disk are also moving around black hole ( synchrotron emission from the accretion) which is depending on the neighborhood of the black hole. Here, to present visual representations of the shadow, we study simple model where there is a free-fall of the accretion onto the black hole from infinity \cite{Jaroszynski:1997bw,Bambi:2012tg,Bambi:2013nla}.
We consider  $a=0$ case for simplification and check the preliminary effect of parameter $L$ on the luminosity of the shadow.

To calculate the specific intensity observed at the observed photon frequency $\nu_{obs}$, we use the following integrating along the light ray:
        \begin{equation}
            I(\nu_{obs},b_\gamma) = \int_\gamma g^3 j(\nu_e) dl_{prop},
            \label{eq:bambiI}
        \end{equation}
where $b_{\gamma}$ is the impact parameter, $\nu_e$  is the photon frequency of the emitter,  $j(\nu_e)$ is the emissivity per unit volume, and $dl_{prop}$ is the infinitesimal proper length. Note that the redshift factor for the infalling accretion is obtained by
        \begin{equation}
            g = \frac{k_\mu u^\mu_o}{k_\mu u^\mu_e}.
        \end{equation}
It is noted that $k^\mu=\dot{x}_\mu$ is the four velocity of the photon, $u^\mu_o=(1,0,0,0)$ is the four velocity of the distant observer. The $u^\mu_e$ is the four velocity of the infalling accretion
     \begin{equation}
u_{\mathrm{e}}^{t}=\frac{1}{A(r)}, \quad u_{\mathrm{e}}^{r}=-\sqrt{\frac{1-A(r)}{A(r) B(r)}}, \quad u_{\mathrm{e}}^{\theta}=u_{\mathrm{e}}^{\phi}=0.
\end{equation}
For the photons, $k_{t}$ is a constant of motion and $k_{r}$ can be inferred from $k_{\alpha} k^{\alpha}=0$ as
    \begin{equation}
k_{r}=\pm k_{t} \sqrt{B(r)\left(\frac{1}{A(r)}-\frac{b^{2}}{r^{2}}\right)}.
\end{equation}
Here the sign plus (minus) stands for the case that the photon gets close to (away from) the black hole. Then using the above relation, one can write the redshift factor $g$ and proper distance $dl_\gamma$ as follows
   \begin{equation}
   g = \Big( u_e^t + \frac{k_r}{k_t}u_e^r \Big)^{-1},
  \end{equation}
  and
 \begin{equation}
  dl_\gamma = k_\mu u^\mu_e d\lambda = \frac{k^t}{g |k_r|}dr.
\end{equation}
 After assuming the monochromatic emission for the specific emissivity with rest-frame frequency $\nu_*$;
        \begin{equation}
            j(\nu_e) \propto \frac{\delta(\nu_e - \nu_*)}{r^2},
        \end{equation}
Then the intensity equation given in \eqref{eq:bambiI} reduces to this form:

        \begin{equation}
            F(b_\gamma) \propto \int_\gamma \frac{g^3}{r^2} \frac{k_e^t}{k_e^r} dr.
        \end{equation}

\begin{figure}[H]
{\centering
\includegraphics[scale=0.5]{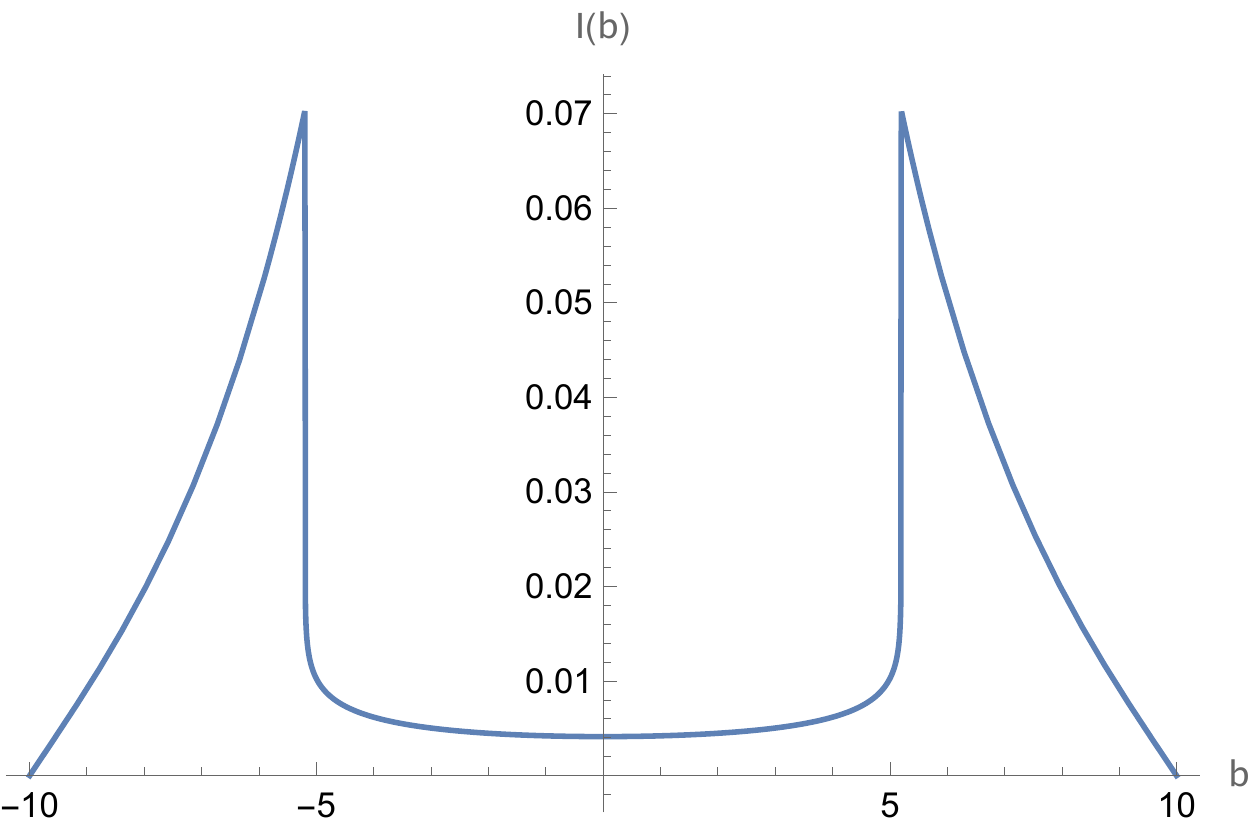}\hspace{1cm}
\includegraphics[scale=0.3]{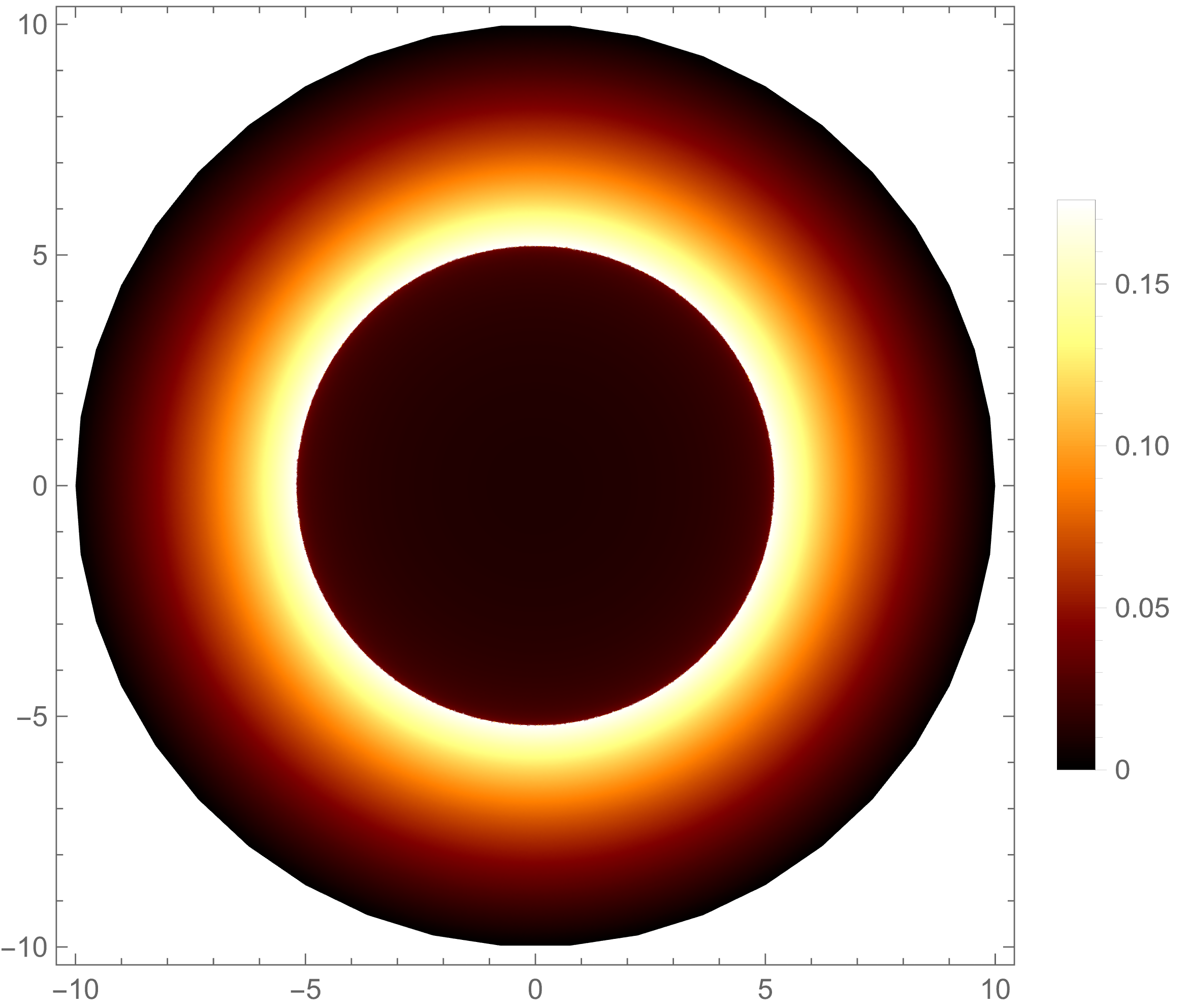}
\caption{The specific intensity $I_{obs}$ seen by a distant observer for an infalling accretion for fixed $L=-0.09$, $M=1$ and $a=0$.}\label{fig:instensity0}	}	
\end{figure}

\begin{figure}[H]
{\centering
\includegraphics[scale=0.5]{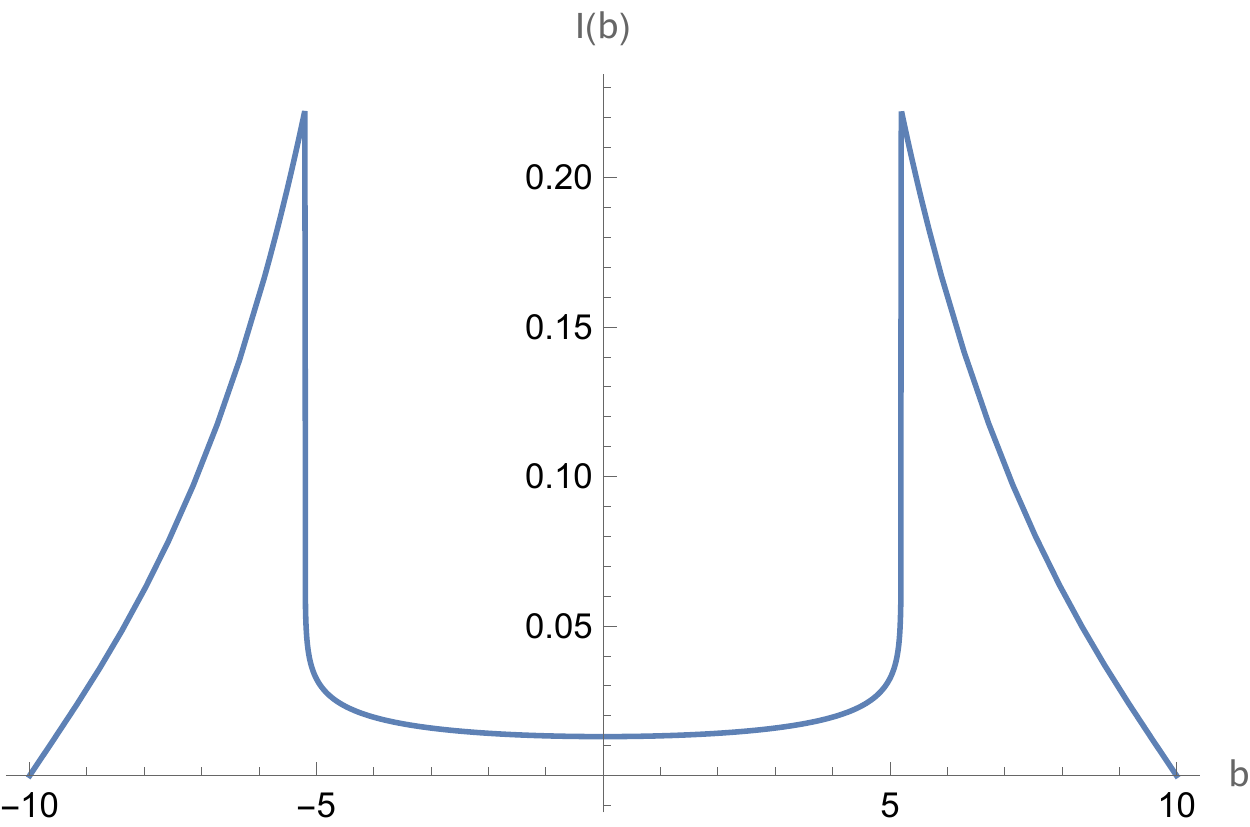}\hspace{1cm}
\includegraphics[scale=0.3]{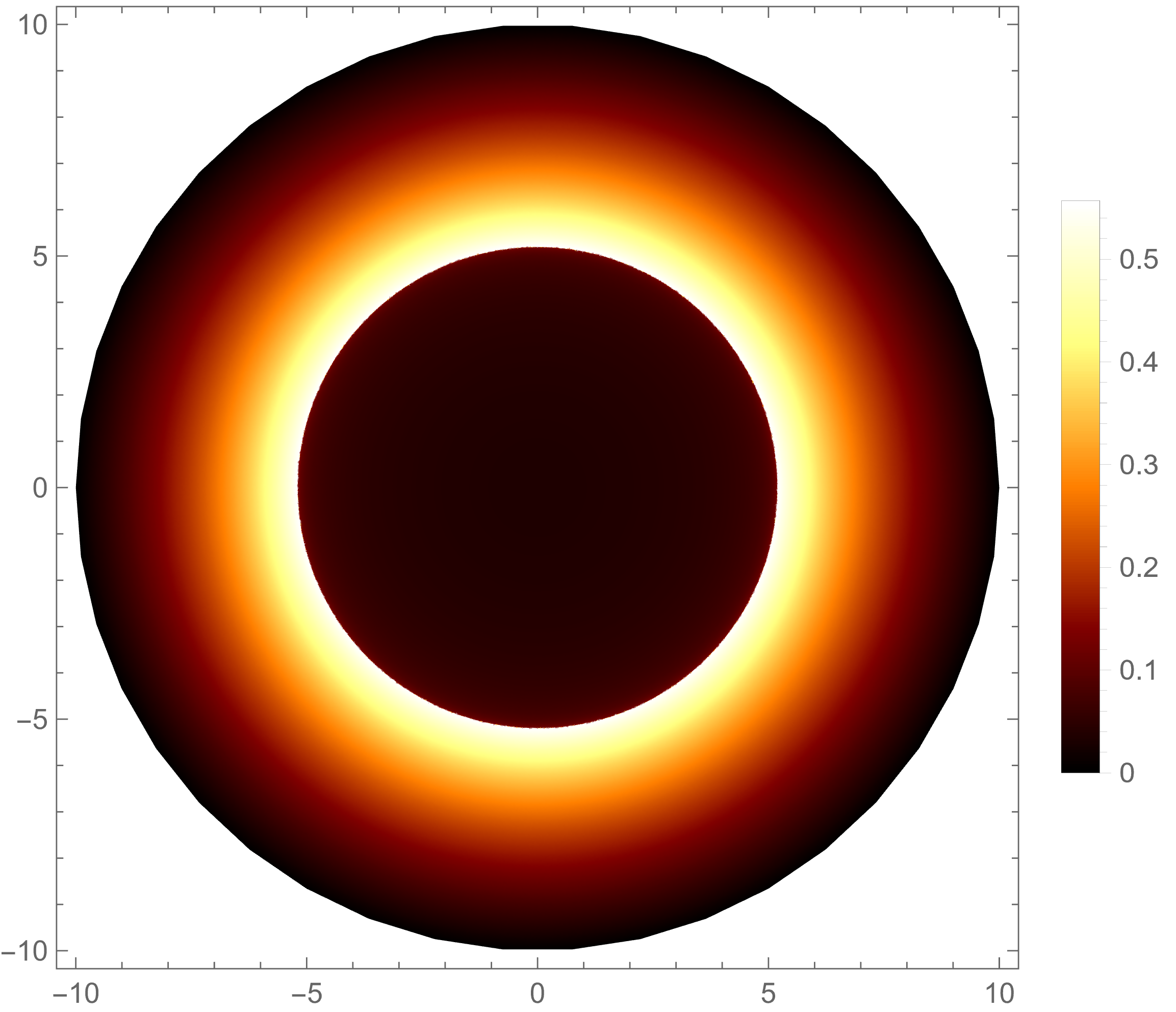}
\caption{The specific intensity $I_{obs}$ seen by a distant observer for an infalling accretion for fixed $L=0$, $M=1$ and $a=0$.}\label{fig:instensity1}	}	
\end{figure}

\begin{figure}[H]
{\centering
\includegraphics[scale=0.5]{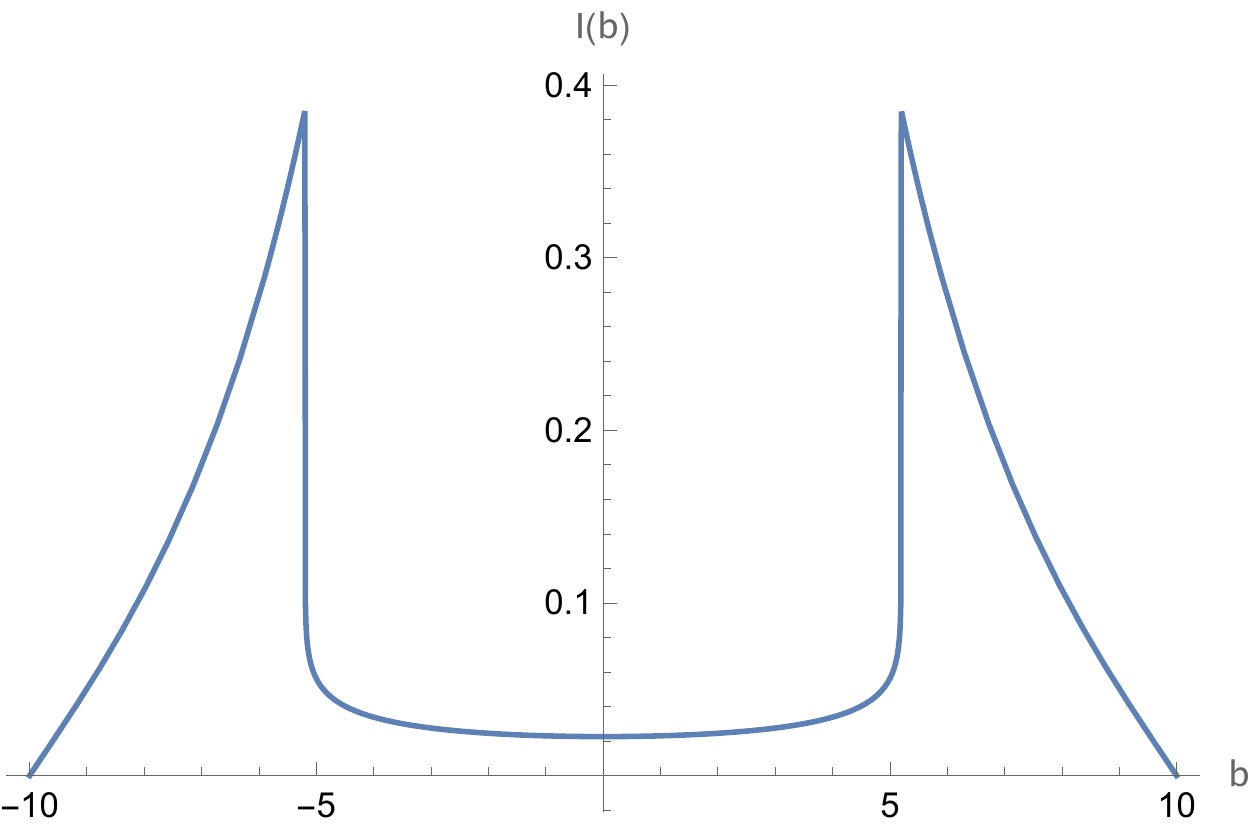}\hspace{1cm}
\includegraphics[scale=0.3]{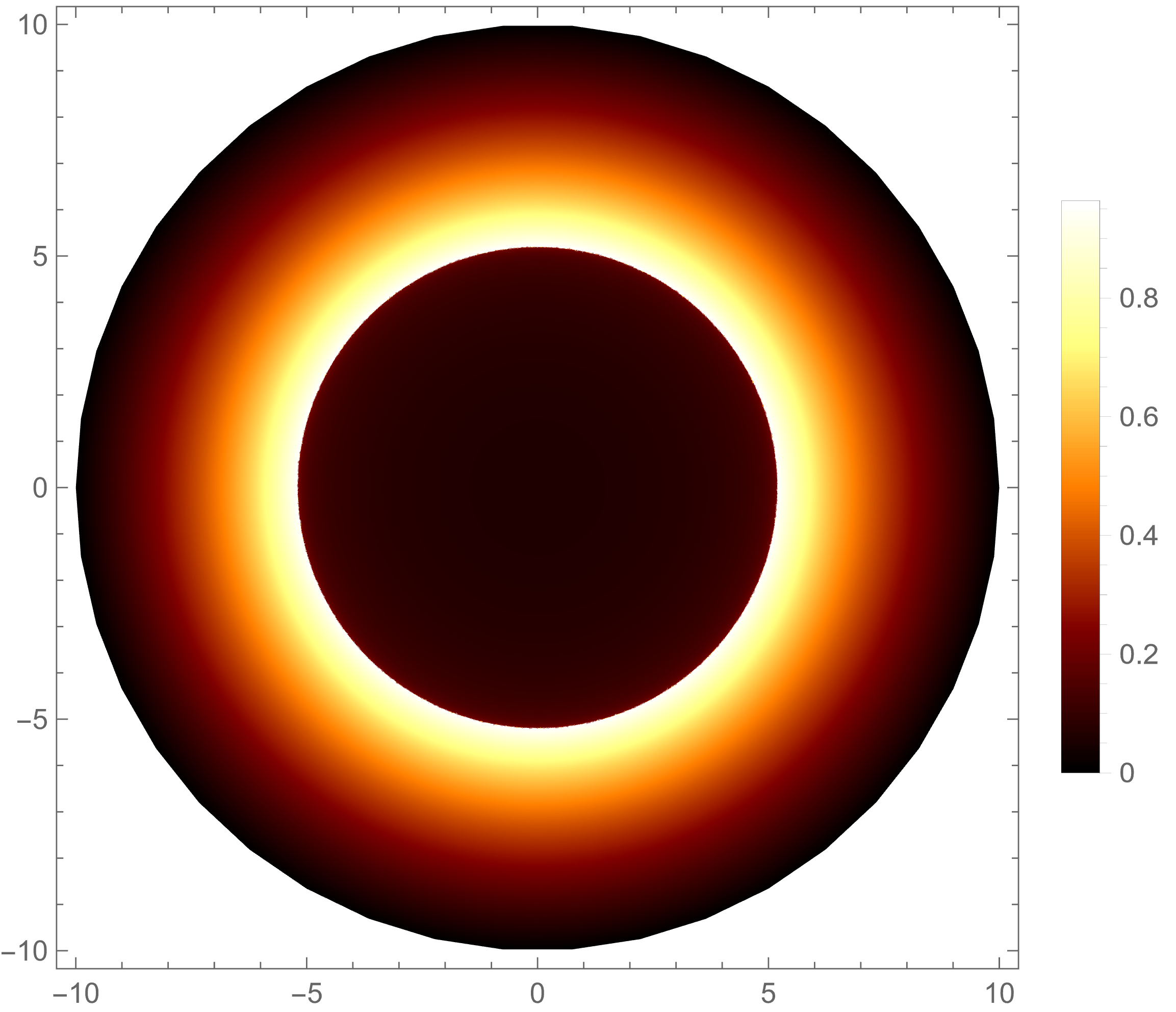}
\caption{The specific intensity $I_{obs}$ seen by a distant observer for an infalling accretion for fixed $L=2$, $M=1$ and $a=0$.}\label{fig:instensity2}	}
\end{figure}

\begin{figure}[H]
{\centering
\includegraphics[scale=0.5]{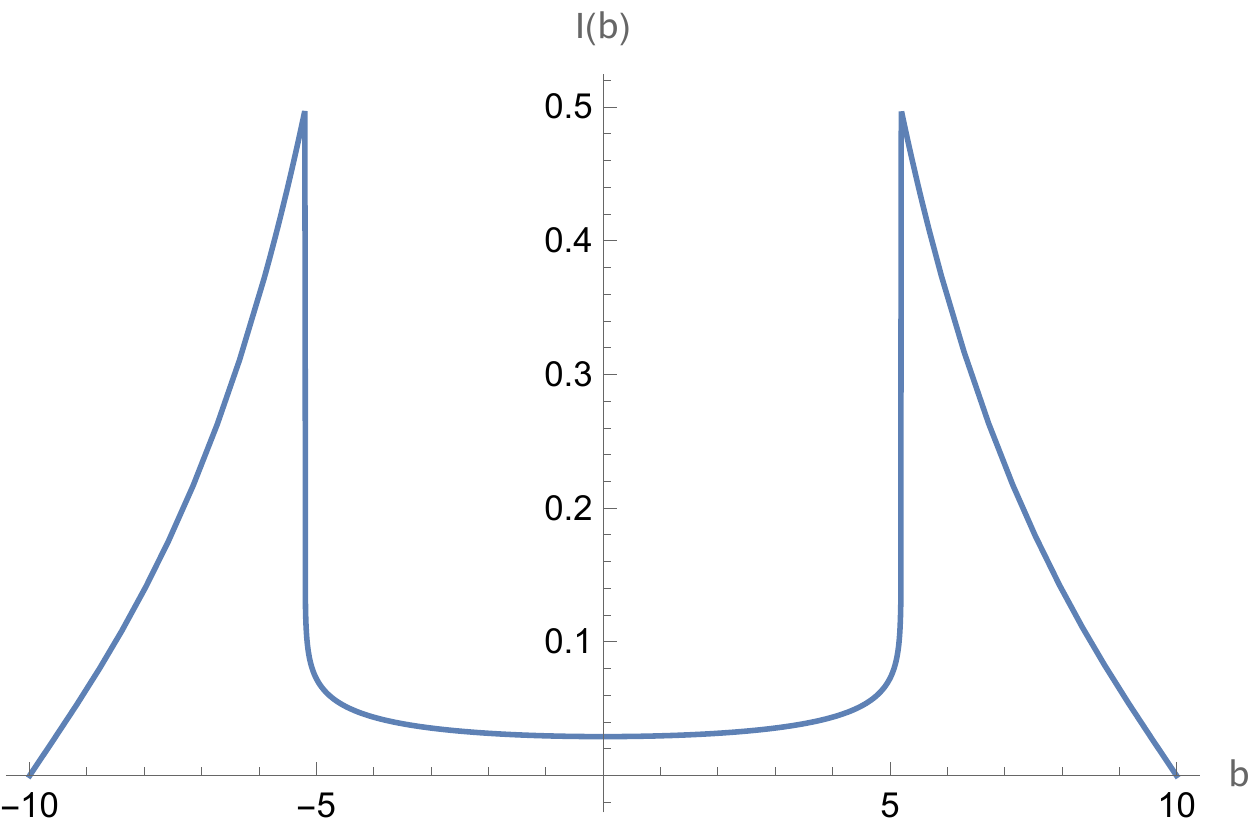}\hspace{1cm}
\includegraphics[scale=0.3]{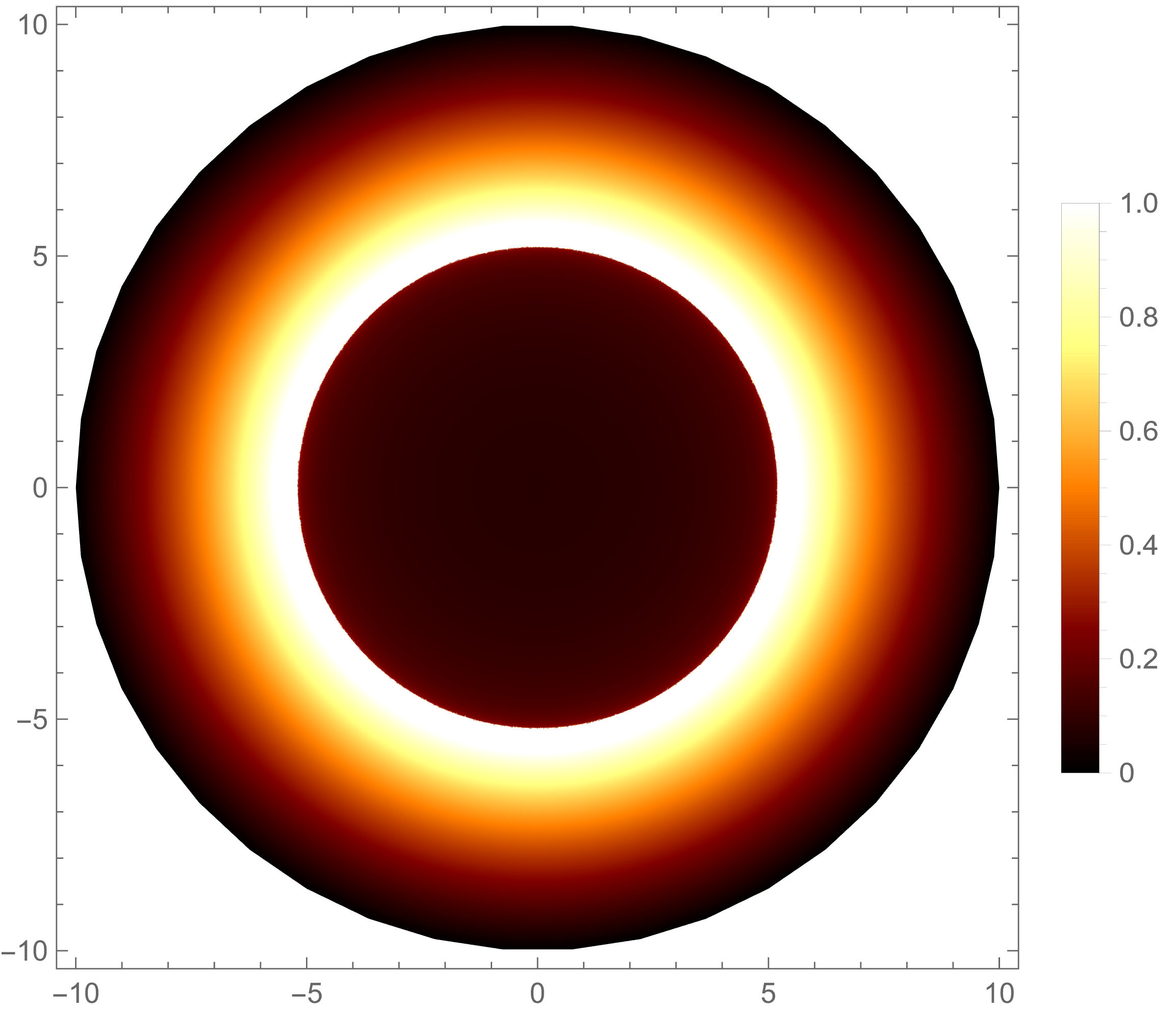}
\caption{The specific intensity $I_{obs}$ seen by a distant observer for an infalling accretion for fixed $L=4$, $M=1$ and $a=0$.}\label{fig:instensity3}	}	
\end{figure}

\begin{figure}[H]
{\centering
\includegraphics[scale=0.5]{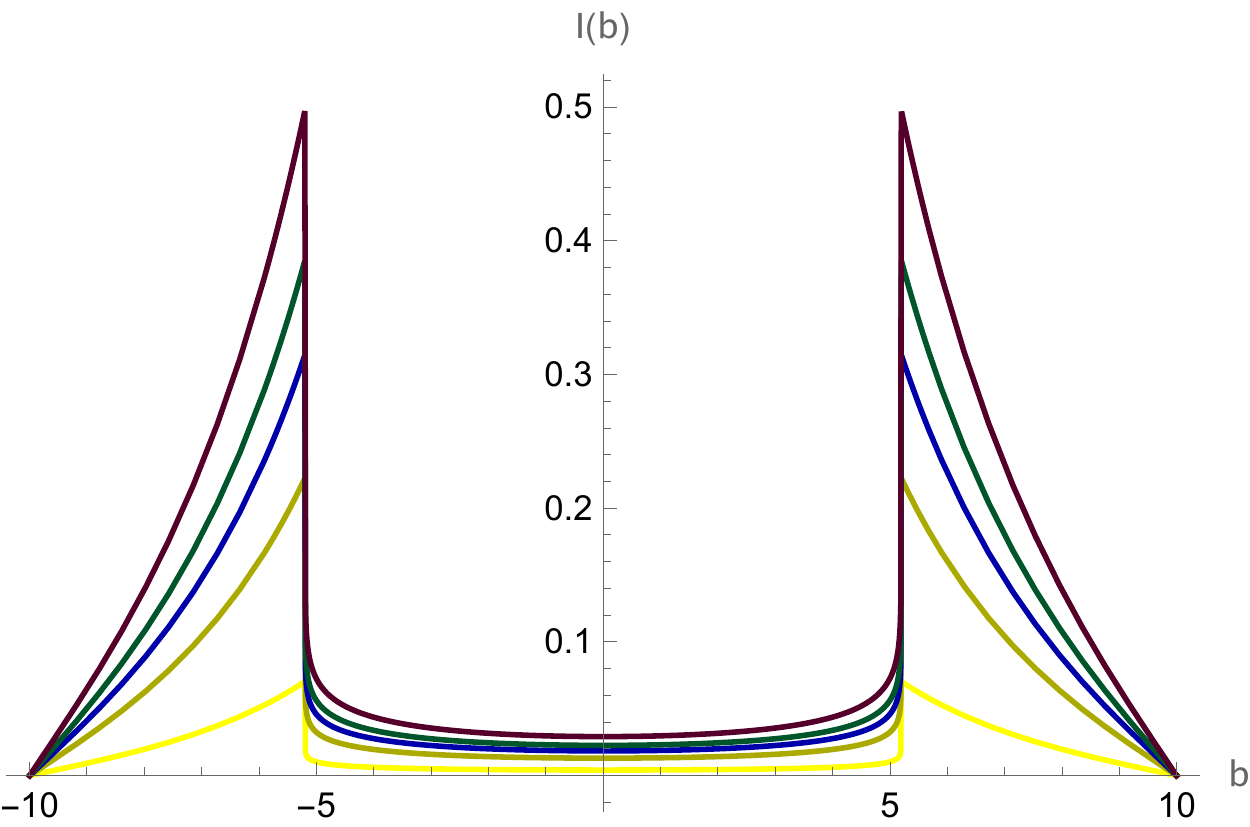}\hspace{1cm}
\caption{The specific intensity $I_{obs}$ seen by a distant observer for an infalling accretion for different $L=-0.09$ (yellow), $L=0$ (orange), $L=1$ (blue), $L=2$ (brown) and $L=4$ (black) with $M=1$ and $a=0$.}\label{fig:instensity4}	}	
\end{figure}

To study the shadow of the black hole with thin-accretion disk, we solve the above equation numerically by using the  \textit{Mathematica} notebook package \cite{Okyay:2021nnh}, (used in \cite{Chakhchi:2022fls}) integrated the flux to see the effects of the parameters of the LV. To this end, we focus on the non-rotating case. For different values of LV parameter $L$, we plot the intensity versus $b$ observed by the distant observer in FIGs. \ref{fig:instensity0}-\ref{fig:instensity3} for instance. We observe that when $b$ increases, intensity getting larger first and then hit the peak value sharply at the photon sphere where the photons are captured by black hole quickly, and then smoothly decreases. Left side of the figures, we show the 2-Dimensional shadow cast image with photon sphere by distant observer in (X, Y) plane. There is a central black region where the event horizon is and this black region is covered by a bright ring (with a photon sphere with the strongest luminosity). After the maximum luminosity of the shadow cast, there is a region of lower brightness which is gradually fading. Note that the luminosity of the shadow increases with increasing the value of the LV parameter $L$. However, there is only a small differences on the intensities of the brightness, To see the clear difference, one should check the Fig. \ref{fig:instensity4}, where we show that increasing the LV parameter $L$, clearly increases the specific intensity seen by a distant observer for an infalling accretion. It is noted that the above analysis is for static case, it is interesting to improve the numeric and study the rotating case and do the comparison with static case, which deserves to be present elsewhere.

\subsection{Shadow  observables and constraint from EHT observations of M87*}

Then to carefully study the effect of LV parameter on the distortion and size of the shadow, it is necessary to connect them with astronomical observables which could be easily measured and helpful to test black hole parameters. Till now, there are different observables which can be  defined via the coordinate of the shadow boundary. We first study two characterized observables, $R_s$ and $\delta_s$ \cite{Hioki:2009na}, where  $R_s$ is the radius of the reference circle for the distorted shadow and $\delta_s$ is the deviation of the left edge of the shadow from the reference circle boundary. In order to give the explicit definitions of the observables, we indicate the top, bottom, right and left of the reference circle as $(X_t, Y_t)$, $(X_b, Y_b)$, $(X_r,0)$ and $(X_l, 0)$, respectively and $(X_l', 0)$ as the leftmost edge of the shadow. Thus, the characterized observables are defined via \cite{Hioki:2009na}
\begin{equation}\label{eq:Rs}
R_s=\frac{(X_t-X_r)^2+Y_t^2}{2\mid X_r-X_t\mid}, ~~~~~   \delta_s=\frac{\mid X_l-X_l'\mid}{R_s},
\end{equation}
and for the shadow of non-rotating case, we have $ \delta_s=0$ because of the shape with perfect circle.

Another two characterized observables,  the shadow area $\mathbb{A}$ and oblateness $\mathbb{D}$, which could describe the shadow with more general shapes,  were proposed in \cite{Kumar:2018ple}. Their definitions are
\begin{eqnarray}
\mathbb{A}=2\int Y{(r_p)}dX{(r_p)}=2\int^{r_p^{max}}_{r_p^{min}}\left(Y{(r_p)}\frac{dX{(r_p)}}{d{r_p}}\right)d{r_p},
~~~~~
\mathbb{D}=\frac{X_r-X_l}{Y_t-Y_b}.
\label{eq-A}
\end{eqnarray}
Note that $\mathbb{D}=1$ for non-rotating case and $\sqrt{3}/{2} \le \mathbb{D}<1$ for Kerr black hole in the view of an equatorial observer, where  $\mathbb{D}=\sqrt{3}/{2}$ is for the extremal case \cite{Tsupko:2017rdo}.

\begin{figure}[H]
{\centering
\includegraphics[scale=0.5]{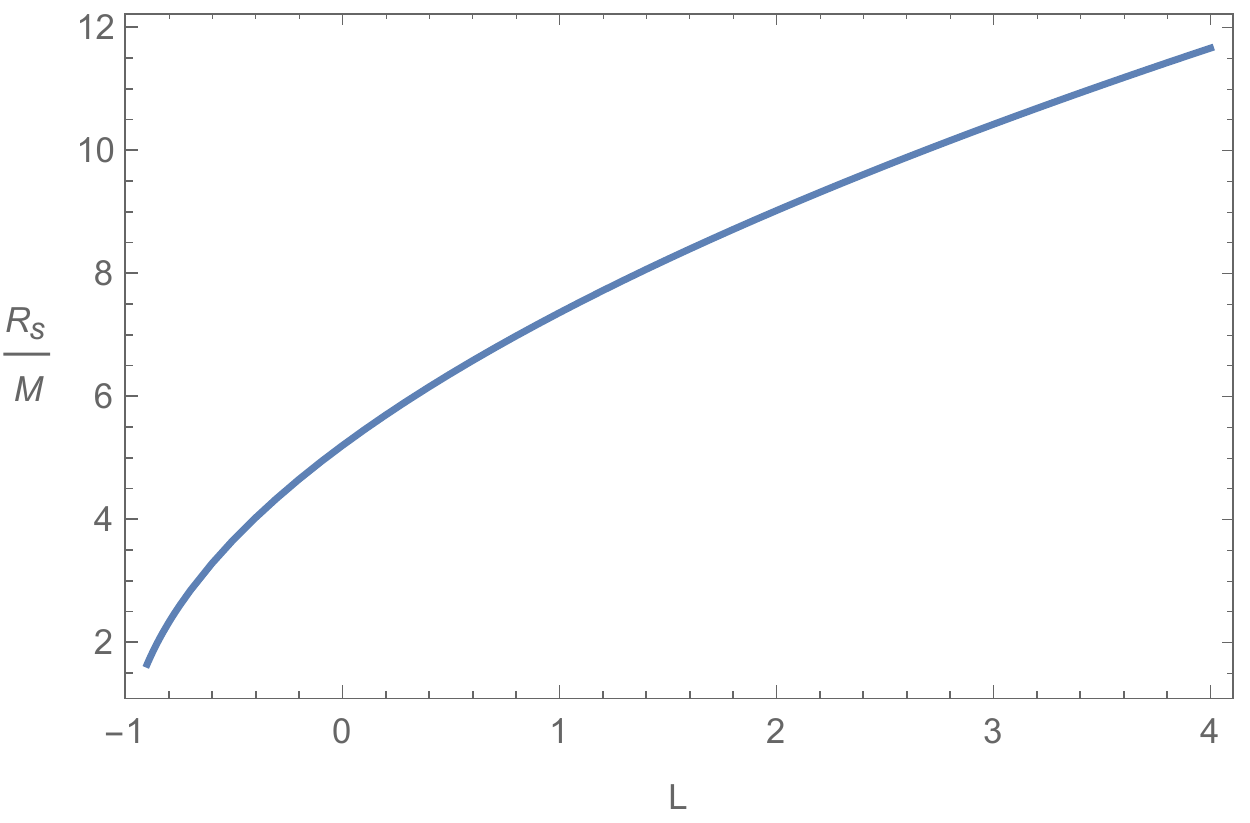}\hspace{1cm}
\includegraphics[scale=0.5]{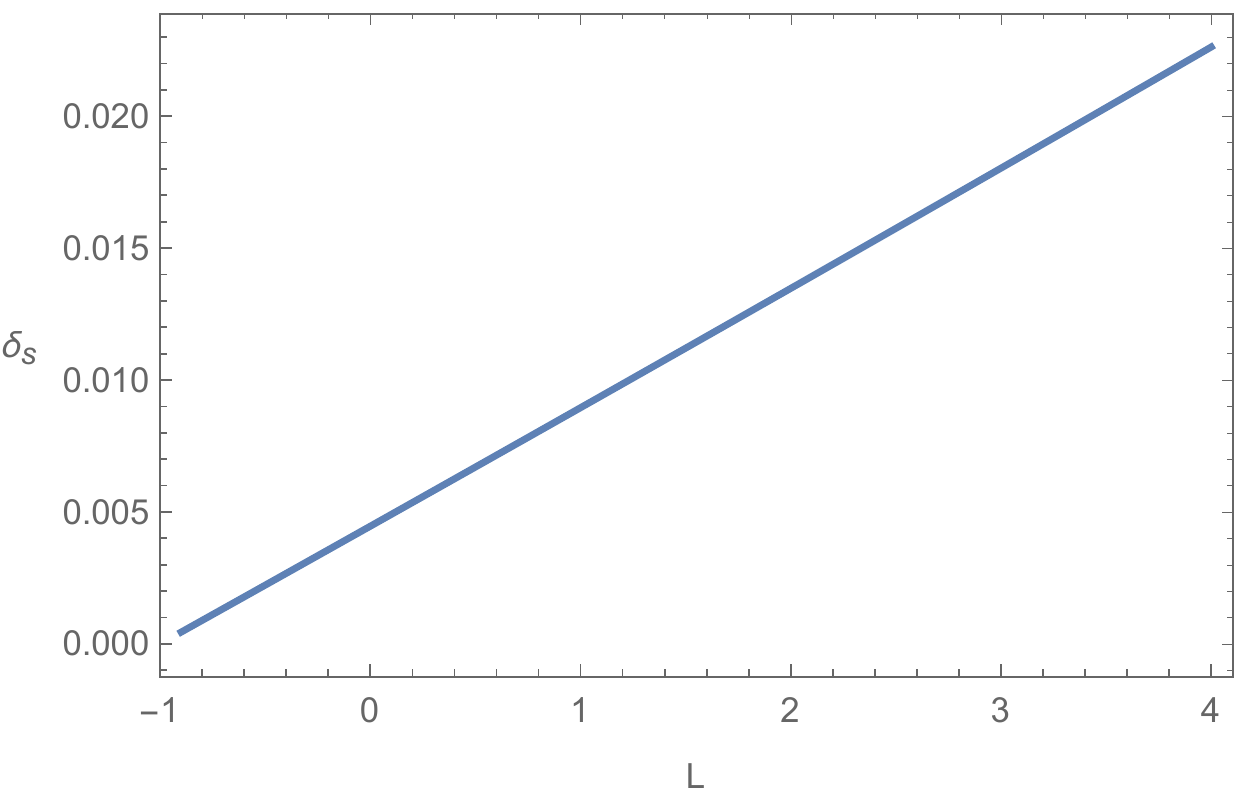}\\
\includegraphics[scale=0.5]{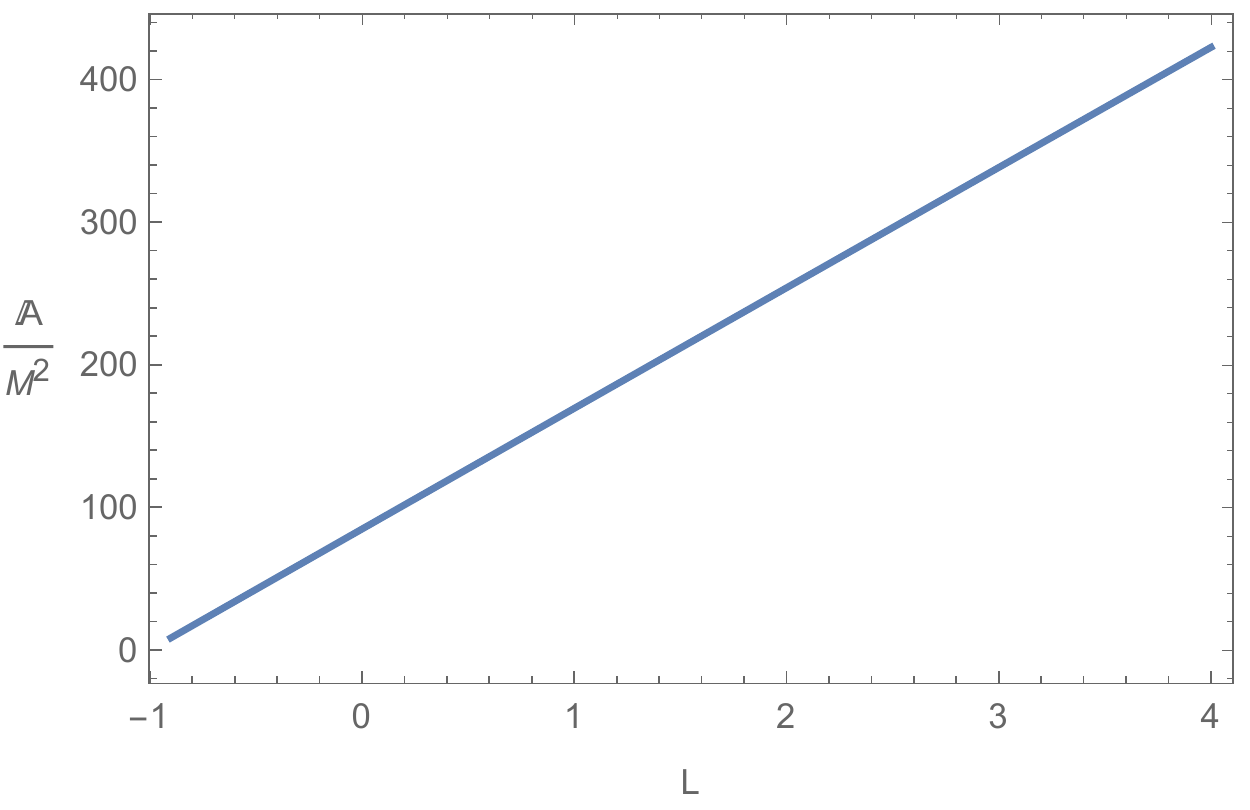}\hspace{1cm}
\includegraphics[scale=0.5]{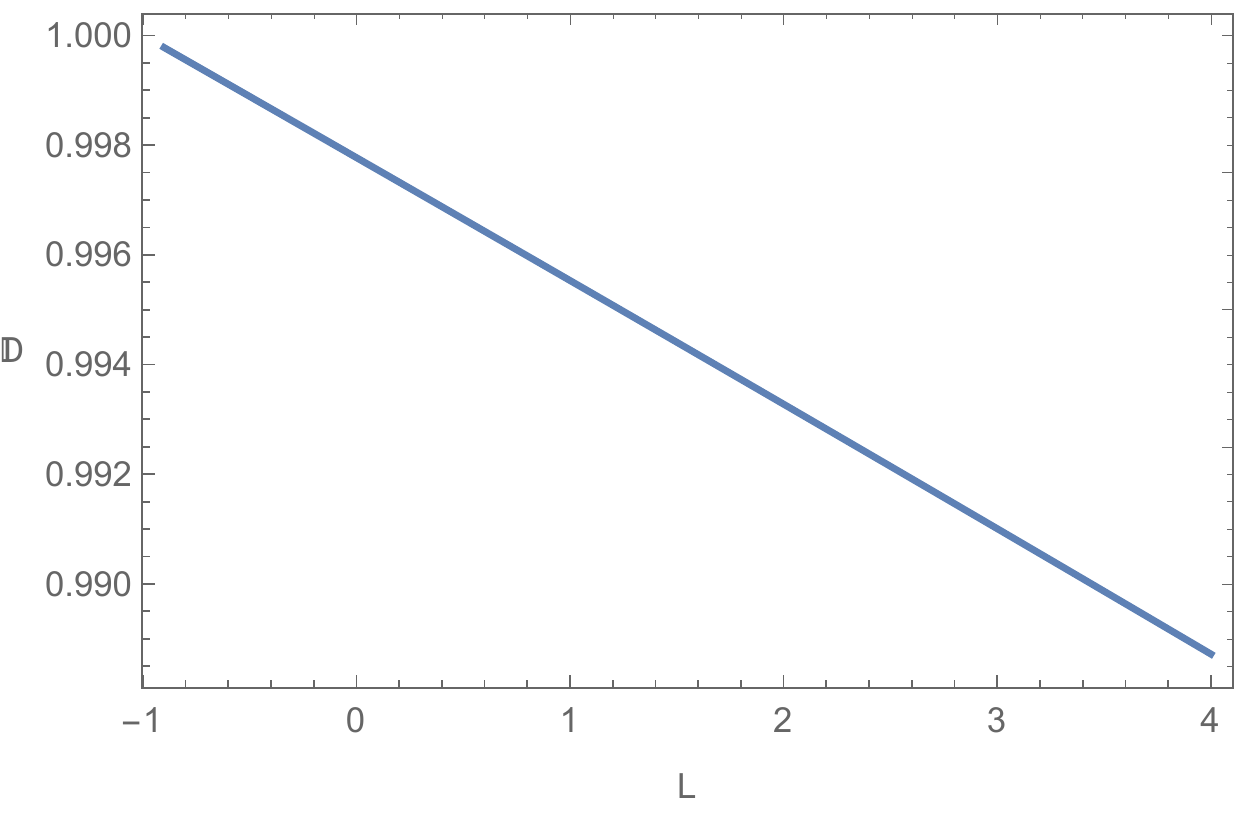}
\caption{Shadow observables, the radius $R_s$ (left upper), the distortion $\delta_s$ (right upper), the area $\mathbb{A}$ (left bottom) and the oblateness $\mathbb{D}$ (right bottom) as functions of $L$ for fixed $a=0.1$ and $\vartheta_o=\pi/2$.}\label{fig:shadowObservables}	}	
\end{figure}

The behaviors of those observables as a function of LV parameter with fixed $a=0.1$ and $\vartheta_o=\pi/2$ are shown in FIG. \ref{fig:shadowObservables}. Then we can read off the following properties: (i) The radius $R_s$ or the area $\mathbb{A}$ increases with the LV parameter. So the shadow of black hole in Einstein bumblebee gravity can be larger (for $L>0$) or smaller ($L<0$) than that of GR black hole. (ii) The distortion $\delta_s$ is non-zero and becomes larger for larger $L$, which means that the shadow of black hole with $L>0$ is more distorted than that of GR black hole, while  the shadow of black hole with $L<0$ is less distorted. (iii) The oblateness $\mathbb{D}$ is smaller than 1 as expected and as $L$ becomes larger, it becomes smaller. So comparing to the GR black hole, the black hole shadow in Einstein bumblebee gravity could be more or less  oblate depending on the sign of the LV parameter.

The above analysis on the characterized observables of shadow provides a possible way to distinguish black holes in GR and Einstein bumblebee gravity. Moreover, if one knows it is a black hole in Einstein bumblebee gravity, then one can determine the sign of the LV parameter by comparing the observables to those of GR black hole.

In \cite{EventHorizonTelescope:2019dse,EventHorizonTelescope:2019ths,EventHorizonTelescope:2019pgp}, it was addressed that the image of supermassive black hole M87* photographed by the EHT is crescent shaped, and the EHT observations constrain the deviation from circularity  as $\Delta C \lesssim 0.1$, the axis ratio as $1<\mathbb{D}_x\lesssim 4/3$ and the angular diameter as $\theta_d=42\pm3 \mu a s$. Using the above EHT constraints, the preliminary analysis of the image of M87* by EHT focus on the Kerr black hole in GR, and the black hole parameters are constrained.
In this subsection, we will consider the M87* black hole as a slowly rotating Kerr-like black hole in
Einstein-bumblebee gravity, and check how the EHT observations to constrain the parameters $a$ and $L$.

To describe the observables related with the EHT observation, circularity deviation $\Delta C$, we recall from last subsection that the distorted black hole shadow is always compared with a reference circle. The geometric center of the black hole shadow can be determined by the edges of the shaped boundary via $({X_c=\frac{X_r+X_l}{2}}, Y_c=0)$. Then the boundary of a black hole shadow can be re-described by the polar coordinates
\begin{equation}
\begin{split}
\phi=\tan^{-1}\left(\frac{Y-Y_C}{X-X_c}\right),~~~~
\mathbb{R}(\phi)=\sqrt{(X-X_c)^2+(Y-Y_c)^2},
\end{split}
\end{equation}
and the average radius of the shadow is defined as
\begin{equation}
\bar{\mathbb{R}}^2=\frac{1}{2\pi}\int^{2\pi}_{0} \mathbb{R}(\phi)^2d\phi.
\end{equation}

As proposed in \cite{Afrin:2021imp}, the circularity deviation $\Delta C$ which measures the deviation from a perfect circle is defined by
\begin{equation}\label{eq-DeltaC}
{\Delta C=\frac{1}{\bar{\mathbb{R}}}\sqrt{\frac{1}{2\pi}\int^{2\pi}_{0}(\mathbb{R}(\phi)-\bar{\mathbb{R}})^2 d\phi}.}
\end{equation}
and the axis ratio is given in \cite{Banerjee:2019nnj}
\begin{equation}\label{eq-Dx}
\mathbb{D}_x=\frac{1}{\mathbb{D}}=\frac{Y_t-Y_b}{X_r-X_l},
\end{equation}
where the  oblateness $\mathbb{D}$ has been defined in \eqref{eq-A}.
Moreover, the angular diameter of the shadow is defined as \cite{Abdujabbarov:2016hnw,Abdujabbarov:2015xqa,Kumar:2020owy}
\begin{equation}\label{eq-theta-d}
\theta_d=2\frac{\mathbb{R}_a}{d}\,.
\end{equation}
Here $\mathbb{R}_a=\sqrt{\frac{\mathbb{A}}{\pi}}$ is the shadow areal radius and $d$ is the distance of the M87* from the earth. It is obvious that the observables related with the EHT observation  depend on the black hole parameters $L$ , $M$ and $a$.

Assuming the supermassive black hole M87* the current slowly rotating Kerr-like LV black hole and set $M=6.5\times 10^{9}M_{\odot}$ and $d=16.8 Mpc$, we could evaluate the observables for the metric \eqref{eq:Kerr-like metric} and use the EHT constrain to  constrain  the parameters $L$ and $a$.  In the calculations, besides setting $\vartheta_o=90^{\circ}$, we shall also consider the case with $\vartheta_o=17^{\circ}$ because as addressed in \cite{CraigWalker:2018vam} the inclination angle (with respect to the line of sight) is estimated to be $\vartheta_o=17^{\circ}$ in the  M87*  if considering the orientation of the relativistic
jets.

\begin{figure}[H]
{\centering
\includegraphics[scale=0.5]{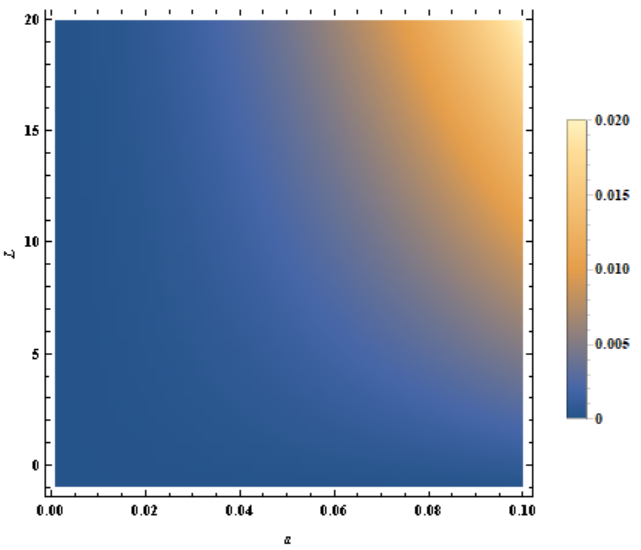}\hspace{1cm}
\includegraphics[scale=0.5]{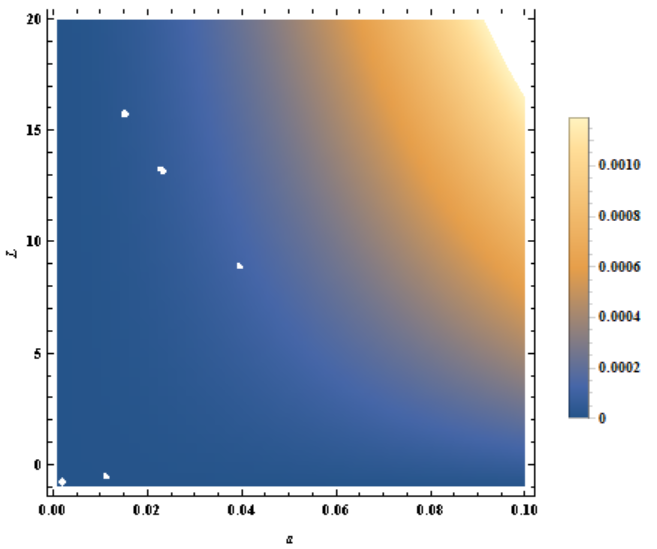}
\caption{The density plot of the circularity deviation $\Delta C$ of the black hole shadow in the $(a-L)$ plane  with $\vartheta_o=90^{\circ}$ (left) and $\vartheta_o=17^{\circ}$ (right). Here we consider the slowly rotating Kerr-like black hole as supermassive black hole in M87*.}\label{fig:DeltaC}	}	
\end{figure}

\begin{figure}[H]
{\centering
\includegraphics[scale=0.5]{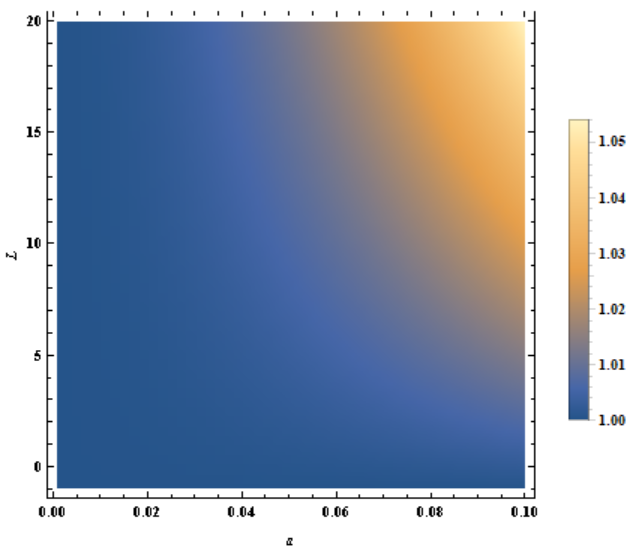}\hspace{1cm}
\includegraphics[scale=0.5]{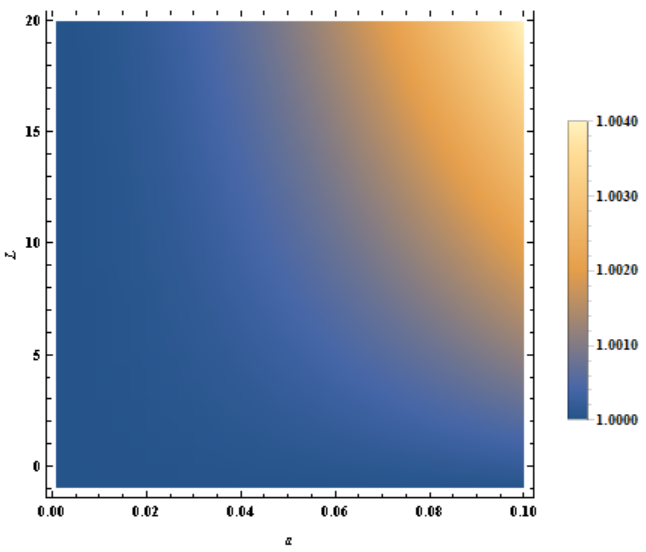}
\caption{The density plot of the axial ratio $\mathbb{D}_x$ of the black hole shadow in the $(a-L)$ plane  with $\vartheta_o=90^{\circ}$ (left) and $\vartheta_o=17^{\circ}$ (right). Here we consider the slowly rotating Kerr-like black hole as supermassive black hole in M87*. }\label{fig:Dx}	}	
\end{figure}

\begin{figure}[H]
{\centering
\includegraphics[scale=0.5]{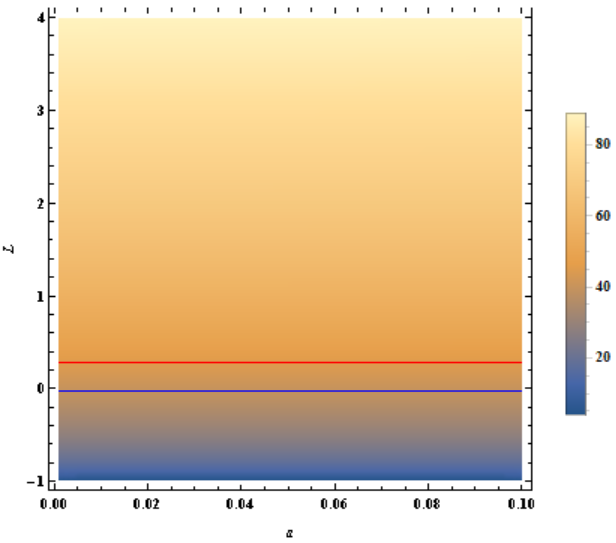}\hspace{1cm}
\includegraphics[scale=0.5]{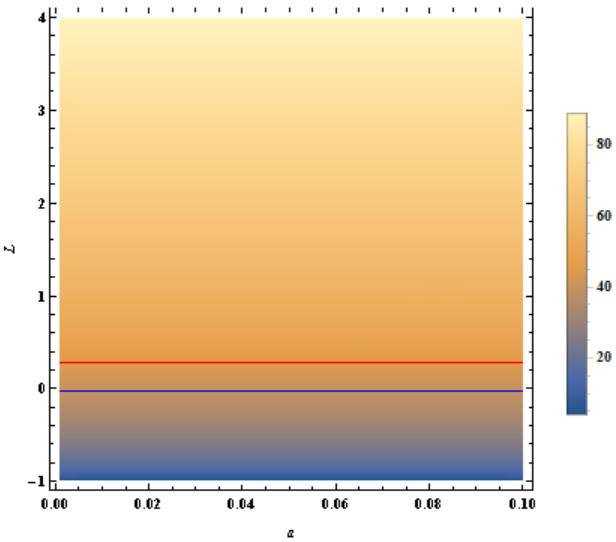}
\caption{The density plot of the angular diameter $\theta_d$ of the black hole shadow in the $(a-L)$ plane  with $\vartheta_o=90^{\circ}$ (left) and $\vartheta_o=17^{\circ}$ (right). Here we consider the slowly rotating Kerr-like black hole as supermassive black hole in M87*. The blue and red curves indicate the $\theta_d=39 \mu as$ and  $\theta_d=45 \mu as$ contours, respectively. } \label{fig:thetad}	}	
\end{figure}

The results in the parameters $(a-L)$ plane are shown in FIG. \ref{fig:DeltaC}-FIG. \ref{fig:thetad}. To explicitly exhibit the phenomena, we confine the range of the spin parameter in $a\in[0,0.1]$ though it could be much smaller in slowly rotating Kerr-like case. FIG. \ref{fig:DeltaC} shows that the EHT observation  $\Delta C \lesssim 0.1$ is satisfied in all the considered parameter region, which implies that it is difficult to use $\Delta C \lesssim 0.1$ in EHT observation to constrain the parameters in slowly rotating Kerr-like black hole \eqref{eq:Kerr-like metric}, or to distinguish Einstein bumblebee gravity from GR.
In FIG. \ref{fig:Dx} of the axial ratio, the EHT observation $1<\mathbb{D}_x\lesssim 4/3$ is also satisfied in the parameter region. This is reasonable because  $\mathbb{D}_x$ is another way to define the circular deviation, and in EHT the emission ring reconstructed is close to circular with an axial ratio of $4:3$ which corresponds to
$\Delta C\lesssim 0.1$ \cite{EventHorizonTelescope:2019dse}. FIG. \ref{fig:thetad} shows the angular diameter $\theta_d$ which is slightly affected by the inclination angle. We see that the EHT observation $\theta_d=42\pm3 \mu a s$ will constrain the black hole parameters in the regime between the blue and red curves, in which the LV parameter can be negative, zero or positive.


\begin{figure}[htp]
   \centering
    \includegraphics[scale=0.3]{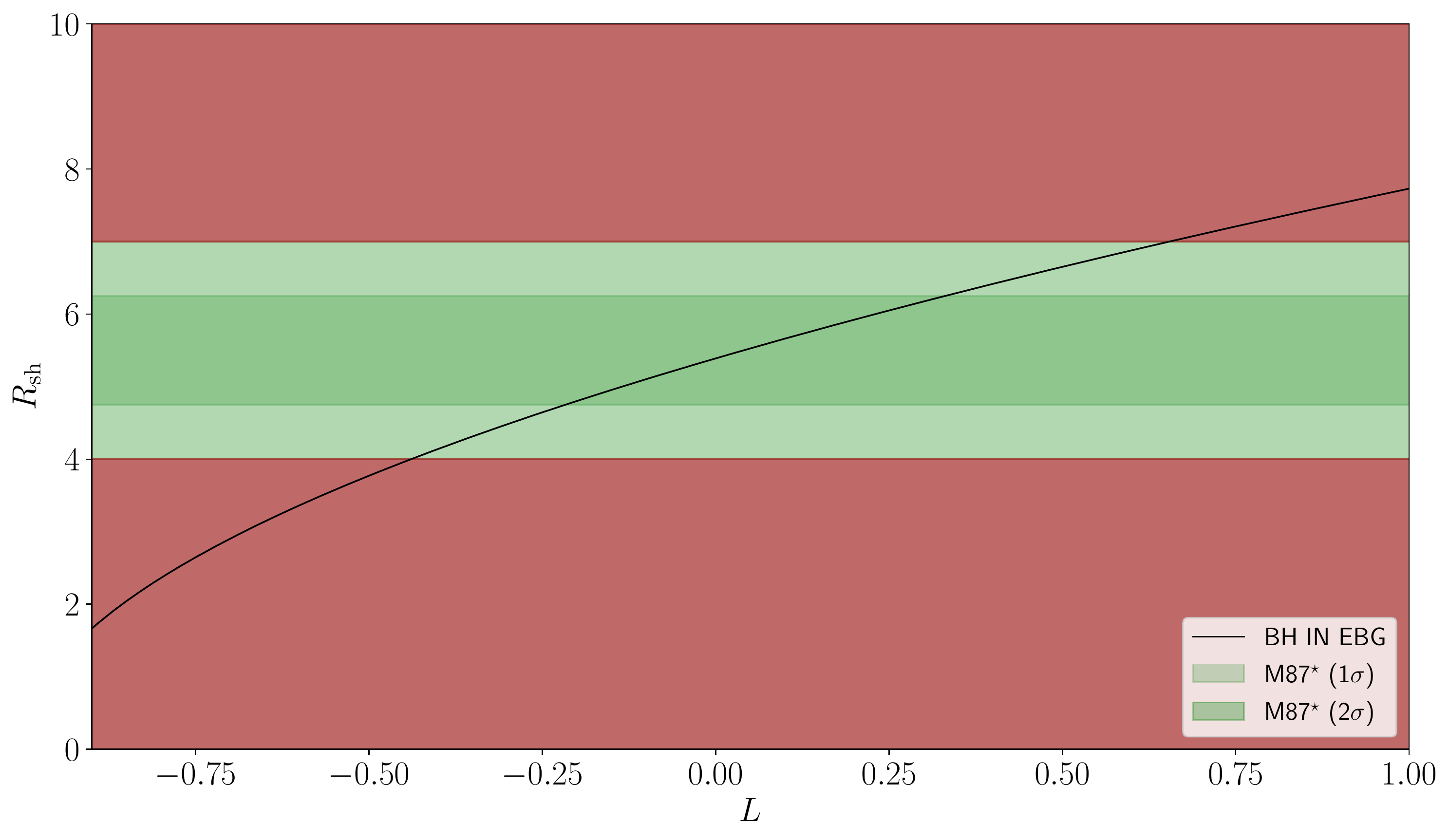}
    \caption{Constraints from EHT  horizon-scale image of M87* at
$1\sigma$ and $2\sigma$ \cite{Vagnozzi:2022moj},  to distance ratio priors for M87* with $a=0.1$ , and varying $L$.}
    \label{fig:shadow.eht}
\end{figure}

Therefore, we can conclude that the EHT observation on M87* black hole shadow cannot rule out the Slowly rotating Kerr-like black hole in Einstein bumblebee gravity. And the current constraints on EHT observation, namely $\theta_d=42\pm3 \mu a s$, provide a possible way to constrain the LV parameter into a smaller range. In particular, since the curves in FIG. \ref{fig:thetad} shows that the spin parameter slightly affects the constraint on the LV parameter, so we fix $a=0.1$ and plot the confidence levels in FIG.\ref{fig:shadow.eht}, which shows that the upper limit of $L$ from the EHT observations that the $68\%$ confidence
level (C.L.) upper limit $L \leq  0.65$, and the $95\%$ C.L.
upper limit  $L \leq 0.35$.

\section{Closing remarks}\label{sec:conclusion}

Gravitational lensing has many applications in solving important astrophysical problems as well as testing GR, and possible alternatives to GR (known as modified gravity theories). Bumblebee gravity is one of them, a very interesting model to explore given the current
interest in theories with LSB. If
it exists, then it will inevitably affect the known relativistic physics especially black hole physics. On the other hand, studying strong
gravitational lensing and also shadow cast of black holes gives us a probe its existence and properties from astronomical observations. \\

In this paper, we have investigated the phenomenology of strong field gravitational lensing by a slowly rotating Kerr-like black hole in Einstein bumblebee gravity. Then we have showed that the effect of LSB causes the different images by shadow of the slowly rotating Kerr-like black hole in Einstein bumblebee gravity and they would be clearly distinguishable and probe the LV. In literature \cite{Ding:2019mal,Jha:2021bue}, some related studies are done on the black hole shadow for the Kerr-like metric, however the solutions does not satisfy the bumblebee field equation, so it may be not proper since the metric is not a solution of the full theory, and the general rotating black hole solution is still an open question for the Einstein bumblebee gravity theory. For this reason, we have used the slowly rotating Kerr-like black hole solution which satisfies the equations of motion. \\

In Section III, we investigate the deflection angle of the strong gravitational lensing, and then evaluate some lensing observables by treating supermassive M87* as lens by the Kerr-like black hole. Firstly, we plot the FIG. \ref{fig:a-xm} to illustrate that when LV parameter increases, the photon sphere increases/decreases for negative/positive value of rotation parameter $a$. Then our result in FIGs. \ref{fig:a-um} and \ref{fig:observables} shows that similarly with an increase of the LV parameter $L$, both the critical impact parameter $u_m$ and the angular position of the relativistic images $\theta_\infty$ increase/decrease with negative/positive values of rotation parameter $a$. Also, the angular separation between the outermost and relativitic images $s$ and the relative magnification $r_{mag}$ are shown to increase/decrease as $L$, and the deviation from the Kerr case depends on the spin parameter.  Moreover, in FIG. \ref{fig:u-alpha} the effect of the LV parameter $L$ is shown on the deflection angle, which should be positive (which happens for only positive $L$) and decreasing with increasing the impact parameter $u$. Lastly, as seen the results in TABLE \ref{table01} and in TABLE \ref{table02}, we have calculated the time delay of the first image from that of the second image, $\Delta T_{2,1}$ for the slowly rotating Kerr-like black hole and then the time delay $\Delta \widetilde{T}_{1,1}$ between prograde and retrograde images  of the same order. Hence, both $\Delta T_{2,1}$ and $\Delta \widetilde{T}_{1,1}$ could be shorter/longer many hours as compared with Kerr case.  which is possible in the current astrophysical observation if one can separate the related two images which could give us hints to distinguish Einstein bumblebee theory from GR. \\

In the Section IV, the observables of the slowly rotating Kerr-like black hole shadows which characterize the shape and distortion as well as the the Lorentz violation parameter constraints from the EHT observation of black hole M87* have been studied. To do so, we have first calculated the null geodesics using the Hamilton-Jacobi method and have obtained the photon sphere of the black hole. First, the shadow radius of the black hole for various LV parameters have been studied in FIG. \ref{fig:shadow}, where we has showed that shape is almost a circle and its size increases with $L$. Moreover, we have showed that the LV parameter $L$ has some positive impact on the luminosity of the photon sphere for the infalling spherical accretion model. The luminosity of the photon sphere around the black hole horizon increases gradually with increased LV parameter $L$ shown in FIGs. (\ref{fig:instensity0}, \ref{fig:instensity1}, \ref{fig:instensity2}, \ref{fig:instensity3} and \ref{fig:instensity4}).  Afterwards, we have showed the observables as a function of LV parameter in FIG. \ref{fig:shadowObservables}. We have concluded that the increasing LV parameter (if $L>0$) provides larger radius $R_s$ as well as distortion $\delta_s$ of the shadow, hence the shadow is more distorted than that of GR black hole. On the other hand, the oblateness $\mathbb{D}$ of the black hole getting smaller for increasing values of LV parameter $L$. Dependence of the LV parameter have been shown clearly and could be observed in the future experiments.  \\

Considering the shadow cast of the slowly rotating Kerr-like black hole solution, we reach the results in the parameters $(a-L)$ plane are shown in FIG. \ref{fig:DeltaC}-FIG. \ref{fig:thetad}. First we have showed in FIG. \ref{fig:DeltaC} that the EHT observation  $\Delta C \lesssim 0.1$ is satisfied in all the considered parameter region, so that it is not possible to distinguish Einstein bumblebee gravity from GR using $\Delta C$. Then in FIG. \ref{fig:Dx} of the axial ratio, the EHT observation $1<\mathbb{D}_x\lesssim 4/3$ is shown that satisfied in the parameter region. Lastly, in FIG. \ref{fig:thetad}, it has been shown that the angular diameter $\theta_d$ which is somewhat deviated by the inclination angle and the EHT observation $\theta_d=42\pm3 \mu a s$ could be constrained the black hole parameters in the regime between the blue and red curves, in which the LV parameter can be negative, zero or positive. As a conclusion, significant deviations may not be detected by measuring the EHT constraints, the preliminary analysis of the image of M87* and it is difficult to use EHT observation to rule out the Slowly rotating Kerr-like black hole in Einstein bumblebee gravity. On the other hand, the recent constraints on EHT observation, namely $\theta_d=42\pm3 \mu a s$, support the way to obtain the LV parameter into a smaller range. \\

 Our results point out that the future generations of EHT observation may help to distinguish the Einstein bumblebee gravity from GR, and also give a possible constrain on the LV parameter, however, recent ones do not allow to safely conclude the LV. Note that in this work we focus on the supermassive M87* black hole from EHT, it would be straightforward to evaluate the lensing or shadow  observables  in our study by supermassive SgrA* black hole.

\begin{acknowledgments}
This work is partly supported by Fok Ying Tung Education Foundation under Grant No. 171006 and Natural Science Foundation of Jiangsu Province under Grant No.BK20211601. A. {\"O}. would like to acknowledge the contribution of the COST Action CA18108 - Quantum gravity phenomenology in the multi-messenger approach (QG-MM).
\end{acknowledgments}

\end{document}